\documentclass[usenatbib]{mn2e}
\usepackage{graphicx,times}
\usepackage{bm}
\usepackage{epsf}
\usepackage{amsmath}
\usepackage{amssymb}
\date{\today,~ $ $Revision: 0.9 $ $}
\def\la{\langle}
\def\ra{\rangle}
\def\n{\noindent}
\def\be{\begin{equation}}
\def\ee{\end{equation}}
\def\ben{\begin{eqnarray}}
\def\een{\end{eqnarray}}
\def\nn{\nonumber}
\def\oh{\hat\Omega}

\def\inc{{\int_0^{r_s}}}

\def\rW{{\rm W}}

\def\bl{{\bf l}}

\def\2p{{(2\pi)^2}}

\def\bl{{\bf l}}
\def\be{\begin{equation}}
\def\ee{\end{equation}}
\def\beq{\begin{equation}}
\def\eeq{\end{equation}}
\def\ben{\begin{eqnarray}}
\def\een{\end{eqnarray}}

\def\oh{{\hat\Omega}}
\def\nn{{\nonumber}}

\newcommand{\beqa}{\begin{eqnarray}}
\newcommand{\eeqa}{\end{eqnarray}}

\newcommand{\rA}{{\rm A}}

\def\exnew{{\cal J}_{\theta_0}(r)}
\def\ikap0{{\cal J}_{\theta_0}(r)}
\def\ikapp{{\cal J}_{\theta_{12}}(r)}

\def\avi{\bar\xi_2^{(i)\kappa}(\theta_0)}

\def\kminnew{{\kappa^{\rm min}_{(i)}}}
\def\one1{\langle \kappa_{(i)}\kappa_{(j)} \rangle}
\def\one{{[\bar \xi^{(ij)}]}}
\def\onei{{[\bar \xi_2^{(i)\kappa}]}}
\def\onej{{[\bar \xi_2^{(j)\kappa}]}}
\def\two{\langle \kappa_{\rm }(\oh_1) \kappa_{\rm }(\oh_2) \rangle_c}
\def\two{{[\xi^{(ij)\kappa}_{12}]}}
\def\corr{{{\cal J}_{\theta_{12}}(r)}}
\def\var{{{\cal J}_{\theta_0}(r)}}

\begin{document}
\onecolumn
\title[Tomography and Weak Lensing Statistics]
{Tomography and Weak lensing Statistics}
\author[Munshi et al.]
{Dipak Munshi$^{1}$, Peter Coles$^{1}$, Martin Kilbinger$^{2,3}$\\
$^{1}$School of Physics and Astronomy, Cardiff University, Queen's
Buildings, 5 The Parade, Cardiff, CF24 3AA, UK \\
$^{2}$Excellence Cluster Universe, Boltzmannstrasse 2, D-85748 Garching bei M\"unchen, Germany \\
$^{3}$Universit\"ats-Sternwarte M\"unchen, Scheinerstr. 1, 81679 M\"unchen, Germany}
\maketitle
\begin{abstract} 
Extending previous studies,
we derive generic predictions for lower order cumulants and their 
correlators for individual tomographic bins as well as between two different bins. We derive the 
corresponding one- and two-point joint probability distribution function for the tomographic
convergence maps from different bins as a function of angular smoothing scale.
The modelling of weak lensing statistics is obtained by adopting a detailed prescription for the underlying density contrast.
In this paper we concentrate on the convergence field $\kappa$ and use top-hat filter; 
though the techniques presented can readily be extended to model the PDF of 
shear components or to include other windows such as the compensated filter. The functional form for the 
underlying PDF and bias is modelled in terms of the non-linear or the quasilinear form
depending on the smoothing angular scale. Results from other semi-analytical models e.g. the lognormal distribution are also presented.
Introducing a {\em reduced} convergence for individual bins, we are able to show that the tomographic PDFs and
bias for each bin sample the same functional form of the underlying PDF of density contrast but with varying variance.
The joint probability distribution of the convergence maps that correspond to two different tomographic bins 
can  be constructed from individual tomographic PDF and {\em bias}.
We study their dependence on cosmological parameters
for source distributions corresponding to the realistic surveys such as LSST and DES.  
We briefly outline how photometric  redshift information can be incorporated in our computation of cumulants, 
cumulant correlators and the PDFs. Various  approximate results for cumulants and their correlators are
presented. Connection of our results to the full 3D calculations is elucidated. Analytical results for inclusion
of realistic noise and finite survey size are presented in detail. 
\end{abstract}
\section{Introduction}
Following the first weak lensing measurements \citep{BRE00,Wittman00,KWL00,Waerbeke00} ,
the field of weak lensing has witnessed a tremendous progress in all fronts (see \citet{MuPhysRep08} for a review).
Currently, in terms of cosmological observations, weak lensing plays a role complementary to both Cosmic Microwave 
Background (CMB) studies and studies involving large scale structure (LSS) surveys.
The ability of weak gravitational lensing to reveal cosmological information, particularly the dark energy equation
of state is considerably enhanced by the inclusion of tomographic information.
The impotance of weak lensing has spurred tremendous progress on the technical front in terms of specification
and control of systematics. There are many ongoing and future weak lensing surveys such as the 
CFHT{\footnote{http://www.cfht.hawai.edu/Sciences/CFHLS/}}
legacy survey, the Pan-STARRS{\footnote{http://pan-starrs.ifa.hawai.edu/}} 
and the Dark Energy survey{\footnote{https://www.darkenergysurvey.org/}},
and further in the future, the Large Synoptic Survey Telescope{\footnote{http://www.lsst.org/llst\_home.shtml}},
Joint Dark Energy Mission or JDEM{\footnote{http://jdem.gsfc.nasa.gov/}} that will map the dark matter and dark energy distribution
of the entire sky in unprecedented details. In particular, owing to the large fraction of the sky coverage
and tighter control on systematics as well as dense sampling of source galaxy populations it will be
soon possible to study gravity induced non-Gaussianity with extreme accuracy.
The gravity induced non-Gaussianity is typically probed using real space correlation
functions as well as in the harmonic domain using their harmonic counterparts i.e.
the multispectra (see e.g. \cite{Pen03}). These correlation functions provide 
a set of tools to go beyond the usual power spectrum analysis. The higher-order correlation functions
are important not only to break the parameter degeneracy  inherent in power spectrum analysis
(e.g. between the amplitude of the matter power spectrum $\sigma_8$ and the matter density parameter $\Omega_{\rm M}$)
but also to understand error-estimates of lower-order correlations functions.
Starting with the study of the three-point correlation function \citep{Vil96,JainSeljak97}
higher order statistics of weak lensing shear, convergence or flexions are now well understood
from a theoretical point of view.

The power spectrum of density perturbations remains the most commonly
used statistic in many cosmological studies. Weak lensing surveys probe the non-linear 
regime and are sensitive to non-Gaussianity which can not be 
probed using only the two-point correlation function or its harmonic analog the power spectrum. 
The statistics of shear or convergence probe the statistics of underlying
mass distribution in an unbiased way \citep{JSW00,MuJai01,Mu00,MuJai00,Valageas00,
VaMuBa05,TakadaWhite03,TakadaJain04}, sensitive to 
nonlinear evolution due to gravitational clustering. Various analytical schemes
from perturbative calculations to halo models have been employed to model
the weak lensing statistics \citet{Fry84,Schaeffer84, BerSch92,SzaSza93, SzaSza97, MuBaMeSch99,
MuCoMe99a, MuCoMe99b, MuMeCo99, MuCo00, MuCo02, MuCo03, CooSeth02}).
In addition to studying the statistics in projection they have also been studied in 3D using photometric
redshifts. This approach can further tighten the constraints on e.g. the neutrino mass
as well as the dark energy equation of state \citep{Heav03,HRH00, HKT06, HKV07, Castro05, Kit08}.
Tomographic techniques have also been employed as an intermediate strategy between projected
surveys and 3D mapping \citep{Hu99,TakadaJain04,TakadaJain03,Semboloni08}.

In this paper we extend previous results \citep{JSW00,MuJai01,Mu00,MuJai00,Valageas00}
on projected surveys by analysing the entire one-point PDF and
the two-point PDF with tomographic information.The PDF contains information about
the correlation hierarchy to an arbitrary order; the correlation hierarchy of 
the convergence field is directly related to that of the underlying mass distribution. 
We employ a generating function formalism that relies on {\em hierarchical ansatz} on
smaller angular smoothing scales and on perturbative results on larger scales.
We define a reduced convergence for each bin and show that the different bins 
sample the same underlying PDF and bias functions (to be defined later) for the density contrast.
The entire joint two-point PDFs for different pairs of redshift bins and 
individual PDF for each bins can be constructed from the PDF and the bias associated
with individual bins because the joint PDF is factorisable in terms of the individual
PDFs, bias and cross-correlations among various bins and different angular scales.
We will show that individual redshift-resolved tomographic maps can be used
to map out the PDF of the underlying mass distribution for a wide range of 
variance. This underlying PDF of the density contrast can be used to 
recover the tomographic PDF with the use of just two individual variables $\kappa_{min}$
and the reduced variance for each bin; both of these variables are uniquely determined by the geometry and 
matter content of the Universe. The results are applicable not only to the PDFs
as determined under hierarchical ansatz but also for other well motivated approximations
for PDF such as the lognormal distribution. 
%

Recent cosmological observations favour an accelerating Universe. This implies existence of
energy of unknown nature (dark energy) which has negative pressure \citep{Amen10,Wang10}. Current data 
continues to be consistent with dark energy being a non-zero cosmological constant.
Though many other alternative dark energy candidates have been consider which are
consistent  with data as well, e.g. quinessence, k-essence, spintessence.
Different dark energy models can be classified according to the equation of state of
of the dark energy component $w_{\rm X}$. For quintessence model  $dw_{\rm X}/dz>0$ while 
for k-essence models  $dw_{\rm X}/dz<0$. There are many complimentary probes for dark energy, 
the distance-redshift relation of cosmological standard candles; 
Cosmic Microwave Background Anisotropy; volume redshift relations using galaxy counts;
the evolution of galaxy clustering; weak lensing, etc. The different methods to probe 
dark energy are complementary to other and can provide important consistency check.
Weak lensing surveys are particularly suitable for dark energy studies. All major 
weak lensing surveys has dark energy as their one of prime science driver. We 
will use the techniques developed in this paper to study two different dark energy
model and compare the predictions against those of standard $\Lambda$CDM model.
The methods presented here are complementary to the usual Fisher matrix based approach
that rely on two-point correlation functions or the power spectrum as it includes
non-Gaussian information upto order .

This paper is organised as follows. In \textsection2 we introduce our notation and present some standard results. 
In \textsection3 we link the lower order statistics of weak lensing convergence to that of the underlying density
distribution. In \textsection4 we briefly review the hierarchical ansatz in the context of 
generating function formalism. In \textsection6 we discuss the lognormal model in the context of weak lensing
statistics. In \textsection7 we derive the PDF and bias for various tomographic bins. The results
are quite generic and can be used for arbitrary source redshift distribution. Finally the \textsection8
is left for  discussion of our results. In an appendix we outline how in the context of tomographic binning
the evolution topological estimators such as Minkowski Functionals can be studied using the lognormal distribution.
\section{Notation}
The statistics of the weak lensing convergence $\kappa_{}(\oh)$ 
represents that of the projected density contrast $\delta({\bf x})$ along the line of sight. 
In our analysis we will consider a small patch
of the sky where we can use the plane parallel approximation or small
angle approximation to replace the spherical harmonics by Fourier modes.
The 3-dimensional density contrast $\delta$ 
along the line of sight when projected onto the
sky with the weight function $\omega_{\rm S}(r,r_s)$ gives 
the weak lensing convergence in a direction $\oh$ which we have denoted by $\kappa_{}({\oh})$:
\be
{\rm Single\;\; Source\;Plane:~~}\kappa_{\rm }({\oh}) = \inc {dr}\;
\omega_{\rm S}(r)\;\delta(r,{\oh}); \quad
\omega_{\rm S}(r,r_s) =  {3\over 2} {H_0^2\over c^2}\ 
\Omega_{\rm M} a^{-1} \ {d_\rA(r) d_\rA(r_s - r)\over d_\rA(r_s)} ; \quad
 \kappa_{\rm S}^{\rm min}(r_s) = -\int_0^{r_s}\omega_{\rm S}(r,r_s) dr.
\ee
Here $d_A(r)$ is the angular diameter distance at a comoving
distance $r$. The subscript $_{\rm S}$ in $\omega_{\rm S}(r,r_s)$ refers to a single source plane.
We have also introduced a parameter $\kappa^{\rm min}$ which will be 
useful in parametrization the PDF and represents the minimum value of the convergence $\kappa$; 
$H_0$ is the Hubble parameter and $a$ represents the scale factor. The comoving radial distance
is denoted by $r$. For a distribution of sources represented by $p_s(z)$ we can write the 
projected convergence $\kappa(\oh)$ as follows:
\be
\omega_{\rm S}(r,r_s) = {3 \over 2}{H_0^2 \over c^2} {\Omega_{\rm M}}a^{-1}(r)  {1 \over \bar n_g} d_\rA(r)\int_r^{r_{\rm H}} dr_s\;p_s(z)  {dz \over dr_s}
{d_\rA(r-r_s)\over d_{\rA}(r_s)}; \quad\quad   p_s(z) = \bar n_g {z^2 \over 2 z_0^3} \exp(-z/z_0).
\ee
\n
In a tomographic analysis the source population is divided into several redshift bins and each of which is treated 
separately. The contribution from the individual bins are taken into account when computing the cumulants or
the cumulants correlators. It is also possible to compute the cross-covariance of these redshift bins.
The convergence $\kappa_{(i)}(\oh)$ from $i$-th tomographic bin can be expressed as:
\ben
&& {\rm Tomography:~~} \kappa_{(i)}(\oh) = \int_0^{r_{\rm H}} w_{(i)}(r)\delta[r,\oh]; \quad\quad
w_{(i)}(r) = {3 \over 2}{H_0^2 \over c^2} \Omega_{\rm M}{1 \over \bar n_i} a^{-1}(r)\;d_\rA(r)\; \int^{r_{i+1}}_{max\{r,r_i\}}\; dr_s \; p_s(z) {dz \over d r_s} 
{d_A(r_s-r) \over d_\rA(r_s)}
\label{eq:omegai}
\een
\n
The ``bin average" of the source population is denoted by $\bar n_i$ and is defined accordingly $\bar n_i =  \int_{r_i}^{r_{i+1}} dr_s p_s(z) {dz/dr_s}$. 
We will consider different bin sizes and source distributions. To incorporate the photometric redshift error we can write
\be
{\rm Photometric\; Redshift\; Errors:~~} w_{(i)}(r) = r\int_{r_i}^{r_{i+1}} dr' \left [ \sum_{h} p_h(z'|z_h)\right ] {\rm F}_{\rm K}(r',r).
\label{eq:omegai_photo}
\ee
Here ${\rm F}_{\rm K} = [S_{\rm K}(r-r')/S_{\rm K}(r)S_{\rm K}(r')]$ with $S_k(r) = \sinh(r), r, \sin(r)$ 
for open, flat and closed geometries. The probability distribution $p_h$ signifies posterior 
probability distribution of redshift given a photometric redshift of  $z_h$.
In our calculation we will need to define a new variable $\kappa^{\rm min}$ (or $\kappa^{\rm min}_{(i)}$ for tomographic bins) which will be useful later:
\be
\kappa_{\rm S}^{\rm min}(r_s) = \int_0^{r_s}\;dr\; w_{S}(r,r_S); \quad \kappa_{(i)}^{\rm min} = \int_0^{r_{\rm H}}\;dr\; w_{(i)}(r); 
\ee
In evaluation of $\kappa_{(i)}^{min}$ we use the following approximate form for the window $w_i(r)$:
\be
 w_{(i)}(r) \approx \Delta r_s {3 \over 2}{H_0^2 \over c^2}\Omega_{\rm M} {1 \over {\bar n_i}} d_{\rm A}(r) p_s(z(r_i)) \left [{dz \over dr_s} \right ]_{r=r_i}
{d_{\rA}(r_i-r) \over d_{\rA}(r_i)} \approx w(r,r_{(i)})
\ee
Using these results it is easy to see that $\kappa^{min}_{(i)} = \kappa^{min}_{\rm S}(r_i)$.

We will adopt two example survey configurations to make definitive calculation. For DES we will take $z_{0}=0.3$ and for LSST we will take $z_0=0.4$.
The range of source distribution that we consider for each survey is $z_s=0.2-1.6$. The bin-size we take is $\Delta z_s = 0.2$.  
The constant $\bar n_g$ is set by imposing the normalized condition $\int_0^{\infty}\; dz\; p_s(z)=1$. 
For our purpose we have $\bar n_g = 1.2 \times 10^7 \bar n_g'$ ($n_g$ specifies the galaxy number density per square arc-minutes). We will vary $n_g$ from few galaxies per arcmin$^2$ to tens of galaxies per arcmin$^2$.
The noise power spectrum $C_l^{\rm N}$ in terms of the intrinsic ellipticity $\gamma_i^2=0.1$ is expressed as $C_l^{\rm N} = \gamma_i^2/\bar n_g$. 

Next we consider the lower order cumulants for individual bins as well as projected catalogs. These results will be 
eventually be useful for the construction of the entire PDF and bias.
The particular cosmology that we will adopt for numerical study are
specified by the following parameter values: $\Omega_\Lambda = 0.741, h=0.72, \Omega_b = 0.044,
\Omega_{\rm CDM} = 0.215, \Omega_{\rm M} = \Omega_b+\Omega_{\rm CDM}, n_s
= 0.964, w_0 = -1, w_a = 0, \sigma_8 = 0.803, \Omega_\nu = 0$.

In addition to the ordinary $\Lambda$CDM model we will also use two dark energy models in our study. 
The angular diameter distance for a dark energy dominated comsology with dark energy equation of state  $\Omega_{\rm X}$ 
can written as:
\be
d_{\rm A}(z) = cH_0^{-1}\int_0^z\; dz'\; [ \Omega_{\rm M}(1+z')^3 + \Omega_{\rm K}(1+z')^2 + \Omega_{\rm X} f(z)]^{-1/2}
\ee
Here $\Omega_{\rm X}$ denotes the dark energy component and $\Omega_{\rm K} = 1-\Omega_{\rm M} -\Omega_{\rm X}$. The function $f(z)$ parametrizes 
the time-dependence of the dark energy density and $f(z=0)=1$. For dark energy with
constant equation of state $w_{\rm X} = p_{\rm X}/\rho_{\rm X}$ we have $f(z) = (1+z)^{3(1+\omega_X)}$. The $\Lambda$CDM is
a limiting case when $w_X=-1$ or $f(z)=1$. In general for  an arbitrary dark energy equation of state
can be represented as \citep{WaGa01} $w_{\rm X}(z) = {1 /3}(1+z)f'(z)/f(z)-1$. The popular parametrization is given 
by $w_{\rm X}(z)=w_0 + w_1 z/(1+z)$. We will consider two different models: (i) constant equation of state $w_0=-0.95$; and
(ii) with evolving equation of state $w_{\rm X}(z)=-1 + {z/(1+ z)}$. 

For the computation of the power spectrum we use the scaling ansatz of \citep{PD94}. The ansatz
consists of postulating a non-local mapping $4\pi k^3P(k) = f_{nl}[4\pi\;k_l^3 P_l(k_l)]$
of linear power spectrum  $P_l(k_l)$ at a wavenumber $k_l$ to nonlinear power spectrum $P(k)$
to another wave number $k$. The wave numbers $k$ and $k_l$ are related by an implicit relation $k_l=(1+4\pi k^3 P(k))^{-1/3}k$.  
The functional form for $f_{nl}$ is determined from numerical simulations (see \textsection\ref{sec:logn} for more related discussions).

The evolution of the linear power spectrum in a dark energy dominated model can be charetcerised using a function $g(z)$ 
i.e. $P_l(k,z)= [g(z)/(1+z)]^2 P_l(k,z=0)$. Where $g(z)$ can be expressed as:
\be 
g(z) = {5 \over 2}\Omega_{\rm M}(1+z)E(z)\int_z^{\infty} dz' {1+z' \over [E(z')]^3}; \quad\quad
E(z) = \sqrt{ \Omega_{\rm M}(1+z)^3 + \Omega_{\rm K}(1+z)^2 + \Omega_{\rm X} f(z)}.
\ee
We will use these expressions to compute the variance of smoothed convergence field $\kappa(\theta_0)$ as a function of source redshift
and smoothing radius $\theta_0$. We will use top-hat smoothing window  $W_{\rm TH}(l\theta_0)$ for our study.

\begin{figure}
\begin{center}
{\epsfxsize=10 cm \epsfysize=5 cm
{\epsfbox[27 440 584 709]{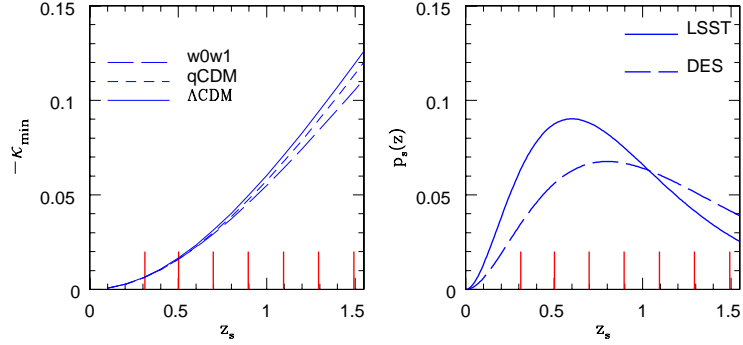}}}
\end{center}
\caption{The parameter $\kappa_{\rm min}$ is plotted as a function of redshift $z_s$ in the left panel for various
background cosmologies. The right panel shows the source density distribution (not normalised) for the two different surveys. The
lines along the x-axis denotes the positions of the tomographic bins considered in our analysis.Notice that
the parameter $k_{\rm min}$ do not depend on smoothing angular scales and only depend on the depth of the survey
as well as on redshift distribution of source population. We consider two different dark energy models along with
$\Lambda$CDM cosmology. The curve qCDM correspond to $w_0=-0.95$ and the model w0w1 correspond to an evolving dark
energy model $w(x)=-1 + z/(1+z)$.}
\label{fig:kmin}
\end{figure}

In Figure (\ref{fig:kmin}) we plot the parameter $\kappa_{\rm min}$ as a function of redshift for different cosmologies (left panel).We also show the number distribution of source galaxies (right panel). 
\section{Lower Order Statistics of Tomographic Convergence Maps}
\label{lower}
Using these definitions we can compute the
projected two-point correlation 
function in terms of the dark matter power spectrum 
$P_{\delta}(k,r)$ (Peebles 1980, Kaiser 1992):
\begin{equation}
\langle \kappa_{(i)}(\oh_1) \kappa_{(j)}(\oh_2) \rangle_c = \inc d {r}
{\omega_{(i)}(r){\omega_{(j)}(r)} \over d^2_A(r)} \int {d^2 {\bf l} \over (2
\pi)^2}~\exp ( i{\bf \theta}_{12} \cdot {\bf l} )~ {\rm P}_{\delta} { \left [ {l\over d_A(r)}, r \right ]}W^2_{\rm TH}(l\theta_0).
\end{equation}
\n
Here $\theta_{12}$ is the angular separation projected onto the surface of the sky 
($\cos|\theta_{12}|=\hat\Omega_1\cdot\hat\Omega_2$) and we have also introduced ${\bf l} = d_A(r){\bf
k}_{\perp}$ to denote the scaled projected wave vector; $\omega_i(r)$ are the weak lensing projection weights for the
ith photometric bins defined in Eq.(\ref{eq:omegai}); to include photometric redshift errors we simply need to use
$\omega_i(r)$ defined in Eq.(\ref{eq:omegai_photo}).
Using Limber's approximation \citep{Limb54} the variance of $\kappa_{(i)}(\theta_0)$
smoothed using a top-hat window $W_{\rm TH}(\theta_0)$ with a radius $\theta_0$ can be written as:
\begin{equation}
\langle \kappa_{(i)}^2 \rangle = \inc d {r}
{\omega_{(i)}^2(r) \over d_A^2(r)} \int {d^2 {\bf l} \over (2
\pi)^2}~ {\rm P}_{\delta} { \Big ( {l\over d_{\rm A}(r)}, r \Big )} W_{\rm TH}^2(l\theta_0).
\label{kappa_variance}
\end{equation}
The variance is plotted for different redshift bins in Figure (\ref{fig:var}).
Similarly the higher order moments of the smoothed temperature
field relate $\langle \kappa^p_{} ({\theta_0}) \rangle$ to the
3-dimensional multi-spectra of the 
underlying dark matter distribution $B_p$ (Hui 1999, Munshi \& Coles 2000):
\beqa
&& \langle \kappa_{(i)}^3 \rangle_c = \inc dr
{\omega_{(i)}^3(r) \over d_{\rm A}^6(r)} \int {d^2 {\bf l}_1 \over (2\pi)^2}
W_{\rm TH}(l_1\theta_0) \int {d^2{\bf
l_2}\over (2\pi)^2} W_{\rm TH}(l_2\theta_0) \int {d^2 {\bf l}_3 \over
(2\pi)^3} W_{\rm TH}(l_3\theta_0) ~ {\rm B}_{\delta} \Big ( {l_1\over d_A(r)},
{l_2\over d_A(r)},  {l_3\over d_{\rm A}(r)}, r \Big )_{\sum {\bf l}_i = 0} \\
&& \langle \kappa_{(i)}^4 \rangle_c = \inc d {r}
{\omega^2_{(i)}(r) \omega^2_{(j)}(r) \over d_{\rm A}^8(r)} \int {d^2 {\bf l}_1 \over (2\pi)^2}
\rW_{\rm TH}(l_1\theta_0) \int {d^2{\bf
l_2}\over (2\pi)^2} \rW_{\rm TH}(l_2\theta_0) \int {d^2 {\bf l}_3 \over
(2\pi)^2} \rW_{\rm TH}(l_3\theta_0)\int {d^2 {\bf l}_4 \over (2 \pi)^2} \rW_{\rm TH}(l_4\theta_0)
\nn \\ &&\quad\quad\quad\quad\quad\quad  ~ \times{\rm T}_{\delta} \Big ( {l_1\over d_A(r)},
{l_2\over d_A(r)}, {l_3\over d_{\rm A}(r)}, {l_4\over d_{\rm A}(r)}, r \Big )_{\sum {\bf l}_i = 0}.
\eeqa
\n
The subscripts $\sum {\bf l}_i = 0$ represent the delta function $\delta_{\rm D}(\sum l_i)$.
We will use these results to show that it is possible
to compute the complete probability distribution function
of $\kappa_{(i)}$ from the underlying dark matter probability
distribution function. Details of the analytical results presented
here can be found in \citep{MuCo00}.
A similar analysis for the higher order cumulant correlators of the
smoothed convergence field relating $\kappa_{(i)}^p (\oh_1) \kappa_{(j)}^q (\oh_2) \rangle_c$ with
multi-spectra of underlying dark matter distribution $B_{p+q}$ can be expressed as \citep{SzaSza97,MuCo00,MuCo02}:
\begin{eqnarray}
&& \langle \kappa_{(i)}^2(\oh_1) \kappa_{(j)}(\oh_2)
\rangle_c =
 \int_0^{r_s} { \omega_{(i)}^2 (r)\omega_{(j)}(r) \over d_{\rm A}^6(r) } dr \int
 \frac{d^2{\bf l}_1}{(2\pi)^2} \int  \frac{d^2{\bf l}_2}{(2\pi)^2} \int \frac{d^2{\bf l}_3}{(2\pi)^2}  \rW_{\rm TH}(
 l_1\theta_0) \rW_{\rm TH}(l_2\theta_0) \rW_{\rm TH}(l_3\theta_0) \nn \\ && \quad\quad\quad\quad \quad\quad\quad\quad \times\exp(i
 {\bf \theta}_{12} \cdot {\bf l}_3){\rm B}_{\delta} \Big ( {l_1\over d_{\rm A}(r)},
{l_2\over d_{\rm A}(r)},  {l_3\over d_{\rm A}(r)}, r \Big )_{\sum {\bf l}_i = 0}.
\end{eqnarray}
We will use and extend these results in this paper to show that it is possible
to compute the whole bias function
$b(>\kappa_{})$, i.e. the bias associated with those spots in 
convergence map which $\kappa_{}$ is above certain threshold (which acts as a generating function for these 
cumulant correlators) from the statistics of underlying over-dense dark objects \citep{MuCoMe99a,MuCoMe99b}. 
\section{Hierarchical {\em Ansatze}}
\begin{figure}
\begin{center}
{\epsfxsize=6 cm \epsfysize=6 cm
{\epsfbox[27 426 316 709]{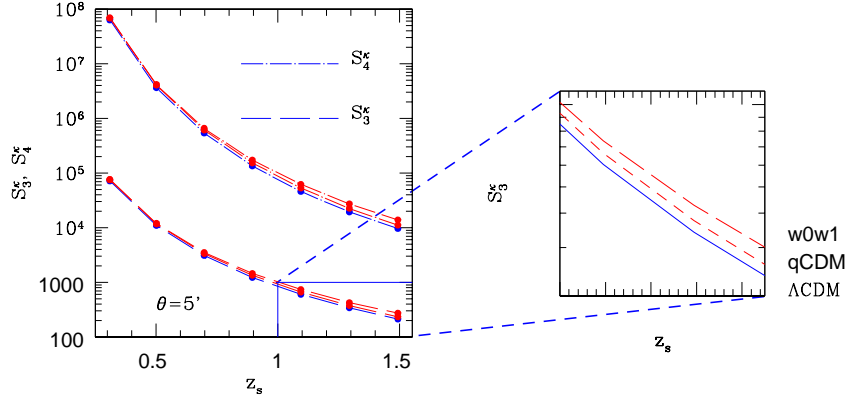}}}
\end{center}
\caption{The skewness parameter $S^{\kappa}_3$ and the kurtosis parameter $S^{\kappa}_4$ is plotted for different redshift bins.
Three different cosmologies are displayed as before.}
\label{fig:sn}
\end{figure}

The spatial length scales corresponding to
small angles are in the highly non-linear regime of gravitational clustering. Assuming a "tree" model 
for the matter correlation
hierarchy in the highly non-linear regime, one can write the 
general form of the $N$th order correlation function $\xi^{(p)}_{\delta}$ as 
(Peebles 1980, Bernardeau \& Schaeffer 1992, Szapudi \& Szalay 1993):
\ben
&& \xi^{(3)}_{\delta}( {\bf r}_1,{\bf r}_2, {\bf r}_3)= Q_3(\xi_{\delta}^{(2)}({\bf r}_1,{\bf r}_2)\xi_{\delta}^{(2)}({\bf r}_1,{\bf r}_3)+ {\rm cyc. perm.} );\\
&& \xi^{(4)}_{\delta}( {\bf r}_1,\cdots, {\bf r}_4)=
 R_a(\xi_{\delta}^{(2)}({\bf r}_1,{\bf r}_2)\xi_{\delta}^{(2)}({\bf r}_1,{\bf r}_3)\xi_{\delta}^{(2)}({\bf r}_1,{\bf r}_4)
+ {\rm cyc. perm.}) + R_b(\xi_{\delta}^{(2)}({\bf r}_1,{\bf r}_2)\xi_{\delta}^{2}({\bf r}_2,{\bf r}_3)\xi_{\delta}^{(2)}({\bf r}_3,{\bf r}_4)
+ {\rm cyc. perm.}).
\een
In general for correlation functions of arbitrary order are constructed by taking a sum over all possible {\em topologies} with
respective amplitudes parameters $Q_{N,\alpha}$, which in general will be different:
\begin{equation}
\xi^{({\rm p})}_{\delta}( {\bf r}_1, \dots {\bf r}_{\rm p} ) = \sum_{\alpha, \rm p-trees}
Q_{p,\alpha} \sum_{\rm labellings} \prod_{\rm edges}^{(\rm p-1)}
\xi^{(2)}_{\delta}({\bf r}_i, {\bf r}_j) .
\end{equation}
To simplify the notation we will use $\xi^{(2)}_{\delta}(r_1,r_2) \equiv \xi_{12}$ and $\bar\xi_2$ for its 
volume average over a volume $v$. It is interesting to note that a similar hierarchy 
develops in the quasi-linear regime in the limit of vanishing variance
(Bernardeau 1992); however the hierarchical amplitudes $Q_{p, \alpha}$
become shape dependent functions in the quasilinear regime. In the 
highly nonlinear 
regime there are some indications that these functions become
independent of shape, as suggested by studies of the
lowest order parameter $Q_3 = Q$ using high resolution numerical
simulations (Sccocimarro et al. 1998). In Fourier space such an
ansatz means that the hierarchy of multi-spectra 
can be written as sums of products of the matter power-spectrum:
\begin{eqnarray}
&&{\rm B}^{(3)}_{\delta}({\bf k}_1, {\bf k}_2, {\bf k}_3)_{\sum k_i = 0} = Q_3 ( P_{\delta}({\bf
k}_1)P_{\delta}({\bf k}_2) + P_{\delta}({\bf k}_2)P_{\delta}({\bf k}_3)
+ P_{\delta}({\bf k}_3)P_{\delta}({\bf k}_1) ) ;\\ 
&&{\rm B}^{(4)}_{\delta}({\bf k}_1, {\bf k}_2, {\bf k}_3, {\bf k}_4)_{\sum k_i = 0} = R_a
\ P_{\delta}({\bf k}_1)P_{\delta}({\bf k}_1 +
{\bf k}_2) P_{\delta}({\bf k}_1 + {\bf k}_2 + {\bf k}_3)  + {\rm cyc. perm.} + R_b(\ P_{\delta}(
{\bf k}_1)P_{\delta}({\bf k}_2)P_{\delta}({\bf k}_3) + 
{\rm cyc. perm.}
\end{eqnarray}
In general for p-the order poly-spectra $B^{(p)}_{\delta}( {\bf k_1}, \dots {\bf k_p} )$ we can write:
\begin{equation}
{\rm B}^{({\rm p})}_{\delta}( {\bf k}_1, \dots {\bf k}_{\rm p} ) = \sum_{\alpha, \rm p-trees}
Q_{{\rm p},\alpha} \sum_{\rm labellings} \prod_{\rm edges}^{(p-1)}
P^{}_{\delta}({\bf k}_i, {\bf k}_j) .
\end{equation}
Different hierarchical models differ in the way they predict the
amplitudes of different tree topologies. Bernardeau \&
Schaeffer (1992) considered the case where amplitudes in general are
factorisable, at each order one has a new ``star'' amplitude 
 and higher order ``snake'' and ``hybrid'' amplitudes can
be constructed from lower order ``star'' amplitudes (see Munshi,
Melott \& Coles 1999a,b,c for a detailed description). In models proposed by
Szapudi \& Szalay (1993) it was assumed that all hierarchical amplitudes of any
given order are degenerate. Galaxy surveys
have been used to study these {\em ansatze}. Our goal here is to
show that weak-lensing surveys can also provide valuable information
in this direction, in addition to constraining the matter power-spectra and
background geometry of the universe. We will use the model proposed by 
Bernardeau \& Schaeffer (1992) and its generalization to the 
quasi-linear regime by Bernardeau (1992, 1994) to construct the PDF
of the weak lensing field $\kappa_{(i)}$. We express 
the one-point cumulants as:
\be
\langle \kappa_{(i)}^3\rangle_c = (3Q_3){\cal C}^{(i)}_3[\kappa^2_{\theta_0}] 
= S_3^{(i)} \langle \kappa_{(i)}^2 \rangle_c^2 \label {hui} \quad\quad
\langle \kappa_{(i)}^4\rangle_c = (12R_a + 4
R_b){\cal C}^{(i)}_4[\kappa^3_{\theta_0}] = S_4^{(i)} \langle \kappa_{(i)}^2\rangle_c^3,
\ee
where we have introduced the following notation:
\beqa
&& {\cal C}^{ij}_{p+q}[[\ikap0]^{p+q-2} [\ikapp] ] =
\int_0^{r_s} { \omega_{(i)}^{p}(r)\omega_{(j)}^q(r) \over
d_A^{2(p+q-1)}(r)}[\ikap0]^{p+q-2}[\ikapp] dr; \\
&& [\ikap0] \equiv  \int
 \frac{d^2\bf l}{(2\pi)^2} P_{\delta} \left( {l \over d_{\rm A}(r)} \right)
{\rm W}_{\rm TH}^2(l\theta_0).\quad\quad
[\ikapp] \equiv  \int
 \frac{d^2\bf l}{(2\pi)^2} P_{\delta} \left( {l \over d_{\rm A}(r)} \right)
{\rm W}_{\rm TH}^2(l\theta_0) \exp ( {\bf l} \cdot {\bf \theta_{12}}).
\eeqa
The normalised cumulants for convergence in the i-th bin are denoted by (skewness) $S_3^{(i)}$ and (kurtosis) $S_4^{(i)}$
and are plotted in Figure (\ref{fig:sn}). 
Eq.({\ref {hui}}) was derived by Hui (1998) in the context of weak lensing surveys. He 
showed that his result agrees well with the ray tracing
simulations of Jain, Seljak and White (1998). Later studies 
extended this result to the entire family of two-point statistics such as cumulant correlators 
(Munshi \& Coles 1999, Munshi \& Jain 1999). 
\begin{eqnarray}
\langle \kappa_{(i)}^2(\oh_1) \kappa_{(j)}(\oh_2) \rangle_c & = &
2Q_3 {\cal C}_3^{(ij)} [\ikap0 \ikapp] =
C_{21}^{\eta}{\cal C}_3^{(ij)} [\ikap0 \ikapp] \equiv C_{21}^{(ij)} \langle
\kappa_{(i)}^2 \rangle_c \langle \kappa_{(i)}(\oh_1) \kappa_{(j)}(\oh_2) \rangle_c, \\
\langle \kappa_{(i)}^3(\oh_1) \kappa_{(j)}( \oh_2) \rangle_c & = &
(3R_a + 6 R_b){\cal C}_4^{(ij)} [\ikap0 \ikapp] =
 C_{31}^{\eta}{\cal C}_4^{(ij)} [\ikap0^2 \ikapp]
 \equiv  C_{(31)}^{(ij)} \langle
\kappa_{(i)}^2 \rangle_c^2 \langle \kappa_{(i)}(\oh_1) \kappa_{j}(\oh_2) \rangle_c
,\\  
\langle \kappa_{(i)}^2(\oh_1) \kappa_{(j)}^2(\oh_2) \rangle_c & =
& 4 R_b{\cal C}_4^{(ij)} [\ikap0^2 \ikapp] 
= C_{(22)}^{\eta}{\cal C}_4^{(ij)} [\ikap0^2 \ikapp] 
\equiv  C_{22}^{(ij)} \langle
\kappa_{(i)}^2 \rangle_c \la \kappa_{(j)}^2 \ra_c\langle \kappa_{(i)}(\oh_1) \kappa_{(j)}(\oh_2) \rangle_c ,\\
\langle \kappa_{(i)}^4(\oh_1) \kappa_{(j)}(\oh_2)\rangle_c & = &
(24S_a + 36S_b + 4 S_c){\cal C}_5^{(ij)} [\ikap0^3
\ikapp] = 
C_{41}^{\eta} {\cal C}_5^{(ij)} [\ikap0^3 \ikapp] 
\equiv  C_{41}^{(ij)} \langle
\kappa_{(i)}^2 \rangle_c^3 \langle \kappa_{(i)}(\oh_1) \kappa_{(j)}(\oh_2)
\rangle_c
,\\ \langle \kappa_{(i)}^3(\oh_1) \kappa_{(j)}^2(\oh_2) \rangle_c & = &
 (12S_a + 6 S_b){\cal C}_5^{(ij)}[\ikap0^3 \ikapp] =
C_{32}^{\eta}{\cal C}_5[\ikap0^3 \ikapp]
\equiv  C_{32}^{(ij)} \langle
\kappa_{(i)}^2 \rangle_c^2 \langle
\kappa_{(j)}^2 \rangle_c \langle \kappa_{(i)}(\oh_1) \kappa_{(j)}(\oh_2)
\rangle_c.      
\end{eqnarray}
where $C_{pq}^{\eta}$ denotes the cumulant correlators for the
underlying mass distribution. These results essentially employ the small angle approximation or Limber's approximation \citep{Limb54}
that are routinely used in computation of higher order cumulants in many cosmological contexts.
Other approximations such as the Born approximation that we use have been verified by testing
against simulations.

In a related, but slightly different context, these lower order statistics can also be helpful in probing the pressure bias
as a function of scale for the study of thermal Sunyaev-Zel'dovich (tSZ) effect or its 
cross-correlation
against tomographic weak lensing maps \citep{Mu11b}. The thermal Sunyaev-Zel'dovich effect probes the line-of-sight
integral of electronic pressure fluctuations. Cross-correlating (frequency-cleaned) $y$ maps 
from ongoing CMB experiments such as Planck \footnote{http://www.rssd.esa.int/index.php?project=SP}\citep{PC06} against
the weak lensing tomographic maps can provide a redshift resolved picture of reionization
history of the Universe. The signal-to-noise will however decrease with increasing order
of these statistics. Indeed the study of PDF or bias that we undertake next will essentially
combine information from all orders. 

\begin{figure}
\begin{center}
{\epsfxsize=6 cm \epsfysize=6 cm
{\epsfbox[27 426 316 709]{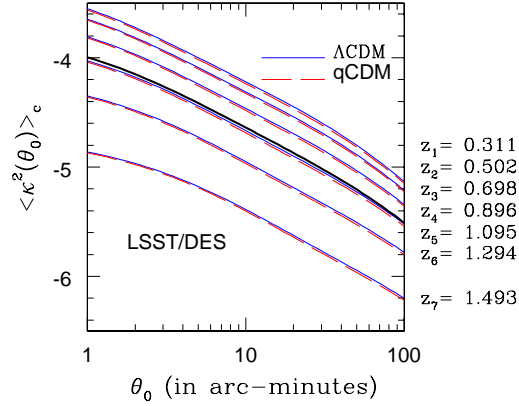}}}
\end{center}
\caption{The plots shows the variance in convergence $\la \kappa^2(\theta_0)\ra_c$ as a function of
smoothing angular scales $\theta_0$. A top-hat window has been assumed. The curves from top to bottom
correspond to various tomographic bins. The redshift bins correspond to $\Delta z_s =0.2$ and covers a range $z_s=0.2-1.4$.The curve qCDM correspond to $w_0=-0.95$.
The higher curves correspond to the deeper redshift bins. See text for more details.}
\label{fig:var}
\end{figure}

In Figure (\ref{fig:var}) and Figure (\ref{fig:sn}) we have plotted the variance and the lower order $S_p$ parameters
respectively as a function of smoothing scale $\theta_0$.
\section{The Generating Function}
\label{gen}
In a scaling analysis of the count probability distribution function (CPDF) the
void probability distribution function (VPF) plays a fundamental 
role. It can be related to the generating function of the cumulants 
or $S_p$ parameters, $\phi(y)$ (White 1979, Balian \& Schaeffer 1989) :
\begin{equation}
P_v(0) = \exp ( -\bar N \sigma(N_c) ) = \exp \Big ( - { \phi (N_c) \over
\bar \xi_2} \Big  ).
\label{eq:scale}
\end{equation}
\n
Where $P_v(0)$ is the probability of having no ``particles'' in a cell of 
of volume $v$, $\bar N$ is the average occupancy of these ``cells'', and 
$N_c = \bar N {\bar \xi}_2$. 
All statistical quantities correspond to underlying density contrast $\delta$. The VPF  $P_v(0)$ is a special case of the count probability
distribution function (CPDF) $P_v(N)$. The VPF $P_v(0)$ contains information of the entire CPDF $P_v(N)$ 

The VPF is meaningful only for a discrete 
distribution of particles and can't be defined for smooth density 
fields such as $\delta$ or $\kappa(\theta_0)$. However the scaling
functions $\sigma(y)$ and $\phi(y)$, defined above in Eq.(\ref{eq:scale}),
$\sigma(y) = -{\phi(y)/ y}$, are very useful even for
continuous distributions where they can be used as a generating
function of one-point cumulants or $S_p$ parameters: $\phi(y) = \sum_{p=1}^{\infty} { S^{\delta}_p/p! } y^p$.
The function $\phi(y)$ satisfies the constraint $S^{\delta}_1 = S^{\delta}_2 = 1$
necessary for proper normalization of PDF. The other generating function
which plays a very important role in such analysis is the generating 
function for the vertex amplitudes $\nu_n$ associated with nodes appearing in the
tree representation of higher order correlation hierarchy ($Q_3 =
\nu_2$, $R_a = \nu_2^2$ and $R_b = \nu_3$).  In practice it is possible
to work with a perturbative expansion of the vertex generating function
${\cal G}(\tau)$. In terms of the vertices this is defined as:
${\cal G}(\tau) =  \sum_{n=0}^{\infty} (-1)^{n} { \nu_n/ n! }$.
However in the highly nonlinear regime a closed form is used.
A more specific model for ${\cal G}(\tau)$, which
is useful to make more specific predictions (Bernardeau \& Schaeffer
1979) is given by ${\cal G}(\tau) = \Big ( 1 + {\tau / \kappa_a} \Big )^{-\kappa_a}$.
We will relate $\kappa_a$ with other parameters of scaling models.
While the definition of VPF does not involve any specific form of
hierarchical {\em ansatz} it is to realize that writing the tree
amplitudes in terms of the weights associated with nodes is only
possible when one assumes a factorisable model of the tree hierarchy
(Bernardeau \& Schaeffer 1992) and other possibilities which do not
violate the tree models are indeed possible too (Bernardeau \&
Schaeffer 1999). The generating functions for tree nodes can be 
related to the VPF by solving a pair of implicit equations 
(Balian \& Schaeffer 1989),
\be
\phi(y) = y {\cal G}(\tau) - { 1 \over 2} y {\tau} { d \over d
\tau} {\cal G}(\tau); \quad\quad \tau = -y { d \over d\tau} {\cal G}(\tau).
\ee
The above description has been limited to the level of constructing
one-point PDF. A more detailed analysis is needed to include the effect of
correlation between two or more correlated volume element which will 
provide information about bias and cumulant correlators. The bias $b(\delta)$
can be introduced the following expression for the joint or two-point PDF:
\be
p(\delta_1,\delta_2)d\delta_1d\delta_2=p(\delta_1)p(\delta_2)(1+ b(\delta_1)\xi_{12}b(\delta_2))d\delta_1d\delta_2
\ee
The function $\tau(y)$ - sometimes denoted by $\beta(y)$ in the literature -
plays the role of a generating function for the
factorized cumulant correlators $C^{\eta}_{p1}$ ($C^{\eta}_{pq} = C^{\eta}_{p1}C^{\eta}_{q1}$) \citep{BerSch92,B92,B94}:
$\tau(y) = \sum_{p=1}^{\infty} y^p {C^{\eta}_{p1}/p!}$.
We will next consider two different regimes; the quasilinear regime valid at large angular scales
and the highly nonlinear regime valid at smaller angular scales.
\subsection {The Highly Non-linear Regime}
\label{subsec:non}bb
The PDF $p(\delta)$ and bias $b(\delta)$  can be related to their
generating functions VPF $\phi(y)$ and $\tau(y)$ respectively
by following equations (Balian \& Schaeffer 1989, Bernardeau \&
Schaeffer 1992, Bernardeau \& Schaeffer 1999),
\be
p(\delta) = \int_{-i\infty}^{i\infty} { dy \over 2 \pi i} \exp \Big [ {(
1 + \delta )y - \phi(y)  \over \bar \xi_2} \Big ]; \quad
b(\delta) p(\delta) = \int_{-i\infty}^{i\infty} { dy \over 2 \pi i} \tau(y) \exp \Big [ {(
1 + \delta )y - \phi(y)  \over \bar \xi_2} \Big ] \label{ber}. 
\ee
The function $\phi(y)$ ($\tau(y)$) plays an important role in any calculation involving hierarchical
ansatz because it completely determines 
the behavior of the PDF $p(\delta)$ (bias $b(\delta)$ ) for all values of $\delta$. The
different asymptotic expressions of $\phi(y)$ govern the behavior
of $p(\delta)$ for different intervals of $\delta$. For large $y$ we
can express $\phi(y)$ as: $\phi(y) = a y^{ 1 - \omega}$.
No theoretical analysis has been done so far to link the newly introduced parameter $\omega$ and the initial power spectral index $n$. In the highly nonlinear regime, numerical simulations are generally
used to fix $\omega$ for a specific initial condition. 
(Colombi et. al. (1992, 1994, 1995), \cite{MuBaMeSch99}). Typically for a power law initial power spectrum 
with pictorial index $n=-2$ one obtains $\omega=0.3$. The VPF
$\phi(y)$ and its two-point analog $\tau(y)$
both exhibit singularity for small but negative value of $y_s$,
\be
\phi(y) = \phi_s - a_s \Gamma(\omega_s) ( y - y_s)^{-\omega_s}; \quad\quad
\tau(y) = \tau_s - b_s ( y - y_s )^{-\omega_s - 1}.
\ee
For the factorisable model of the hierarchical clustering the 
parameter $\omega_s$
takes the value $-3/2$ and $a_s$ and $b_s$ can be expressed in terms
of the  nature of the generating function ${\cal G}(\tau)$ and its 
derivatives near the singularity $\tau_s$ 
(Bernardeau \& Schaeffer 1992):
\be
a_s = {1 \over \Gamma(-1/2)}{\cal G}'(\tau_s) {\cal G}''(\tau_s) \left [
{ 2 {\cal G}'(\tau_s) {\cal G}''(\tau_s) \over {\cal G}'''(\tau_s)}
\right ]^{3/2}; \quad\quad
b_s = \left [
{ 2 {\cal G}'(\tau_s) {\cal G}''(\tau_s) \over {\cal G}'''(\tau_s)}
\right ]^{1/2}.
\ee
As mentioned before the parameter $k_a$ which we have introduced in
the definition of
${\cal G}(\tau)$ can be related to the parameters $a$ and $\omega$ appearing 
in the asymptotic expressions of $\phi(y)$ (Balian \& Schaeffer 1989, 
Bernardeau \& Schaeffer 1992),
\be
\omega = k_a / ( k_a + 2),\label{ka}; \quad\quad
a = {k_a + 2 \over 2} k_a^{ k_a /  k_a + 2}.
\ee
Similarly the parameter $y_s$ which describes the behavior
of the function $\phi(y)$ near its singularity can be 
related to the behavior of
${\cal G(\tau)}$ near $\tau_s$ which is the solution of the equation
(Balian \& Schaeffer 1989, Bernardeau \& Schaeffer 1992),
$\tau_s = {{\cal G}'(\tau_s)/ {\cal G}''(\tau_s) }$,
finally we can relate $k_a$ to $y_s$ by following expression (see eq. (\ref{ka})):
$y_s = - { \tau_s / {\cal G}'(\tau_s)}$, or we can write:
\begin{equation}
-{ 1 \over y_s} = x_{\star} = {1 \over k_a } { (k_a + 2)^{k_a + 2} \over (k_a + 1)^{k_a+1}}.
\end{equation}
The newly introduced variable $x_\star$ will be useful to define the
large $\delta$ tail of the PDF $p(\delta)$ and the bias $b(\delta)$. 
The asymptotes of $\phi(y)$
are linked with the behavior of $p(\delta)$ for various regimes of
$\delta$. For very large values of the variance $\bar\xi_2$ 
it is possible to define a scaling function $p(\delta) = { h(x)/\bar xi_2^2
} $  which will encode 
the scaling behavior of the PDF, where $x$ plays the role of the scaling 
variable and is defined as $x={(1 + \delta)}/\bar \xi_2$. We list below
different ranges of $\delta$ and specify the behavior of $p(\delta)$
and $b(\delta)$ in these regimes (Balian \& Schaeffer 1989).
\begin{equation}
{\bar \xi }^{ - \omega/( 1 - \omega)} \gg 1 + \delta \gg \bar \xi;
~~~~~~
p(\delta) = { a \over \bar \xi_2^2} { 1- \omega \over \Gamma(\omega)}
\Big ( { 1 + \delta \over \xi_2 } \Big )^{\omega - 2}; ~~~~~
 b(\delta) = \left ( {\omega \over 2a } \right )^{1/2} { \Gamma
(\omega) \over \Gamma [ { 1\over 2} ( 1 + \omega ) ] } \left( { 1 +
\delta \over \bar \xi_2} \right)^{(1 - \omega)/2}
\end{equation}
\begin{equation}
1+ \delta \gg {\bar \xi}_2; ~~~~
p(\delta) = { a_s \over \bar \xi_2^2 } \Big ( { 1 + \delta \over \bar
\xi_2}  \Big ) \exp \Big ( - { 1 + \delta \over x_{\star} \bar \xi_2}
\Big );  ~~~~~ b(\delta) = -{ 1 \over {\cal G}'(\tau_s)} {(1 + \delta)
\over { {\bar \xi}_2}} 
\label{eq:bias}
\end{equation}
\begin{figure}
\begin{center}
{\epsfxsize=10. cm \epsfysize=5. cm
{\epsfbox[27 426 590 709]{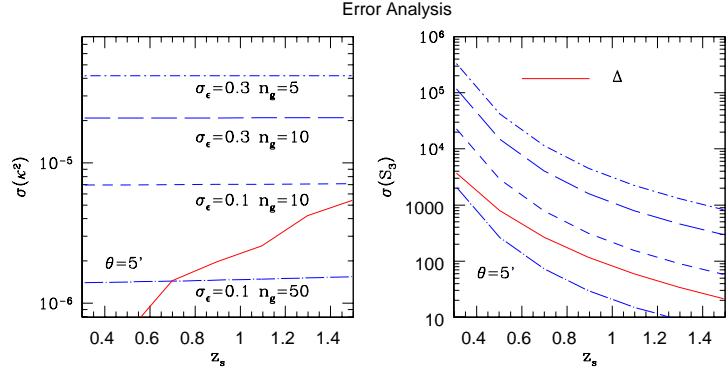}}}
\end{center}
\caption{The left panel shows the $1\sigma$ error in estimation of the variance $\la\kappa^2\ra_{\rm c}$ and
the right panel shows the error of estimation in the skewness paramter $S_3$ as a function of tomographic redshift.  The angular scale
is fixed at $\theta_s=5'$. Various curves correspond to a choice of
the intrinsic ellipticity distribution of galaxies $\sigma_{\epsilon}$ 
and number of galaxies $n_{g}$ (=number of galaxies/arcmin$^2$) are as depicted. The solid lines are the difference between the $\Lambda$CDM model and the
qCDM model. The scatter is computed using the formalism developed in \citep{VaMuBa05}.}
\label{fig:sn_err}
\end{figure}
The integral constraints satisfied by scaling function $h(x)$ are
$ S^{\eta}_1 = \int_0^{\infty} x h(x) dx = 1$ and 
$ S^{\eta}_2 = \int_0^{\infty} x^2 h
(x) dx = 1$. These take care of  
normalization of the function $p(\delta)$. Similarly the 
normalization constraint over $b(\delta)$ or equivalently $b(x)$ can be expressed as
$C^{\eta}_{11} = \int_0^{\infty} x b(x)h(x)dx = 1$, which translates into
$\int_{-1}^{\infty} d\delta b(\delta)p(\delta) = 0$ and
$\int_{-1}^{\infty} d\delta b(\delta)p(\delta) = 1$.  
Several numerical
studies have produced the behavior of $h(x)$ and $b(x)$
for different initial conditions (e.g. Colombi et al. 1992,1994,1995; Munshi et
al. 1999, Valageas et al. 1999). For very small values of $\delta$ the behavior of
$p(\delta)$ is determined by the asymptotic behavior of $\phi(y)$ 
for large values of $y$, and it is possible to define another scaling function 
$g(z)$ which is completely determined by
$\omega$, the scaling parameter can be expressed as $z = (1+
\delta)a^{-1/(1-\omega)}{\bar \xi}_2^{\omega /(1 - \omega)}$. 
However numerically it is much easier to determine $\omega$ 
from the study of $\sigma(y)$ compared to the study of $g(z)$ 
(e.g. Bouchet \& Hernquist 1992).
\begin{equation}
1 + \delta \ll \bar \xi_2;~~~~
p(\delta) = a^{ -1/(1 - \omega)} [{\bar \xi}_2]^{ \omega/(1 -
\omega) } \sqrt { ( 1 - \omega )^{ 1/\omega } \over 2 \pi \omega z^{(1
+ \omega)/ \omega } } \exp \Big [ - \omega \Big ( {z \over 1 - \omega}
\Big )^{- {{1 - \omega} \over \omega}} \Big ]; ~~~~~~~b(\delta) = -
\left ( {2 \omega \over \bar{ \xi}_2} \right )^{1/2} \left ({ 1 -
\omega \over z}  \right )^{(1 - \omega)/2 \omega} 
\end{equation}
To summarize, the entire behaviour of the PDF
 $p(\delta)$ and bias $b(\delta)$ are
encoded in two different scaling functions, $h(x)$ and $g(z)$. 
These scaling functions are relevant for small and large $\delta$ behavior
of the function $p(\delta)$ and $b(\delta)$. Typically the PDF
$p(\delta)$ shows a cutoff at
both large and small values of $\delta$ and it exhibits a
power-law in the middle. The power law behavior is prominent in highly non-linear
regime. With the decrease in $\bar \xi_2$ the range of $\delta$ 
for which $p(\delta)$ shows such a power law behavior decreases
finally to vanish for the case of very small variance i.e. in the
quasi-linear regime. Similarly the bias is a very small and slowly
varying function for moderately over dense objects but increases
rapidly for over-dense objects. These deductions are in qualitative agreement 
with results from the halo model based approaches.
\subsection{The Quasi-linear Regime}
\label{subsec:quasi}
The Generating function formalism was used by \citep{B92,B94} in the quasilinear
regime. Unlike highly nonlinear regime the quasilinear regime can be dealt with 
using perturbative analysis. This particular analysis assumes the variance 
of the smoothed density contrast is smaller than unity. The generating function
formalism was use to construct the PDF and bias using the tree-level
perturbation theory to arbitrary order. In this regime the scaling parameters 
$\omega$ and $k_a$ can be expressed in terms of the initial power spectral index $n$.
In general the numerical values of the parameters $k_a$ or $\omega$ characterizing VPF
or CPDF are different from there highly non-linear values. 
The PDF and bias now can be expressed in terms of $G_{\delta}(\tau)$
(Bernardeau 1992; Bernardeau 1994): 
\begin{eqnarray}
&&p(\delta)d \delta = { 1 \over -{\cal G}_{\delta}'(\tau) } \Big [ { 1 - \tau {\cal G}_{\delta}''(\tau)
/{\cal G}_{\delta}'(\tau) \over 2 \pi {\bar \xi}_{2} }  \Big ]^{1/2} \exp \Big ( -{ \tau^2
\over 2 {\bar \xi}_{2}} \Big ) d \tau; ~~~~~ b(\delta) = - \left (
{k_a \over \bar \xi_2} \right ) \left [ ( 1 + {\cal G}_{\delta}(\tau)
)^{1/k_a} - 1 \right ] , \\
&&{\cal G}_{\delta}(\tau) = {\cal G}(\tau) - 1 =  \delta.
\end{eqnarray}
The above expression is valid for $\delta < \delta_c$ where the $\delta_c$
is the value of $\delta$ which cancels the numerator of the pre-factor 
of the exponential function appearing in the above expression. For
$\delta > \delta_c$ the PDF develops an exponential tail which is 
related to the presence of singularity in $\phi(y)$ in a very similar
way as in the case of its highly non-linear counterpart (Bernardeau
1992; Bernardeau 1994).
\begin{equation}
p(\delta) d \delta = { 3 a_s \sqrt {{\bar \xi}_2} \over 4  {\sqrt \pi} }
\delta^{-5/2} \exp \Big [ -|y_s|{ \delta \over {\bar \xi}_{2}} + {|\phi_s|
\over {\bar \xi}_{2}} \Big ] d \delta; ~~~~~b(\delta) = -{ 1 \over
{\cal G}'(\tau_s)} {(1 + \delta)
\over { {\bar \xi}_2}} 
\end{equation}
These expressions were used by \cite{MuJai01, MuJai00} and \cite{Valageas00} for the construction of 
weak lensing PDF in projection. The bias was studied in \citep{Mu00} for projected surveys.
The tests against numerical simulations show remarkable agreement for a range of angular scales.
Later studies refined these results as well as incorporated various different smoothing windows 
\citep{MuVaBa04,VaMuBa05}. We extend these results derived for surveys in
projection to tomographic surveys in this paper.

It is worth mentioning that,
there have been various attempts to extend the perturbative results to the highly nonlinear
regime (see e.g. \citep{CBBH97,VaMu04})
\begin{figure}
\begin{center}
{\epsfxsize=7. cm \epsfysize=7. cm
{\epsfbox[27 426 316 709]{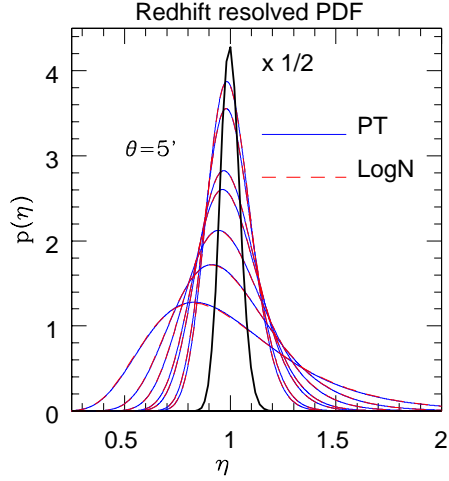}}}
\end{center}
\caption{The PDF $p(\eta)$ of the {\em reduced} convergence $\eta$ as a function $\eta=1+\delta$. The plots
with decreasing peak height correspond to lower redshift bins. Two different approximations
are being compared. The solid line correspond to the lognormal approximation and the
dashed line correspond to the perturbative calculations. The results are shown for a smoothing
angular scale $\theta_s=5'$.}
\label{fig:eta_pdf}
\end{figure}
\section{The Lognormal Distribution}
\label{sec:logn}
An alternative to the hierarchical ansatz, the {\em lognormal} distribution, for the description
of the matter PDF which
has long been known as a successful 
empirical prescription for the characterization of the dark matter distribution as well as the
 observed galaxy distribution \citep{Ham85,CJ91,Bouchet93,Kf94}. Detailed discussion for comparison 
of lognormal distribution and the 
perturbative calculations can be found in \citep{BK95}. The lognormal distribution
was further generalized to the {\it skewed}-lognormal distribution \citep{Col94}. In general a
variable might be modelled as lognormal if it can be thought of as the multiplicative product of 
many independent random variables.

 Although inherently local in nature, the lognormal distribution can provide a good fit to both  
one-point PDF and its generalisation to compute its two-point analog and hence the bias \citep{TTHF02}. 
The one- and two-point lognormal PDF can be expressed as \citep{KTS01}:
\begin{eqnarray}
&& p_{\rm ln}(\delta)d\delta = {1 \over \sqrt {2\pi \bar\Sigma}} \exp \left [ -{\Delta^2 \over 2\Sigma^2}\right ]{d\delta \over 1+\delta}; \quad\quad
\Sigma^2=\ln(1+\sigma^2); \quad\quad \Delta = \ln[(1+\delta)\sqrt{(1+\sigma^2)};\label{eq:logn1a} \\
&& p_{\rm ln}(\delta_1,\delta_2)d\delta_1 d\delta_2 =  {1 \over 2\pi \sqrt {\Sigma^2 - X_{12}^2}}\exp \left [ -{\Sigma(\Delta_1^2 + \Delta_2^2) -2X_{12}\Delta_1\Delta_2 \over 2(\Sigma^2 - X_{12}^2)}\right ] {d\delta_1\over 1+\delta_1} {d\delta_2\over 1+\delta_2}; \label{eq:logn2}\\
&& \Delta_i = \ln[(1+\delta_i)\sqrt{(1+\bar\xi_2^2)};  \quad \Sigma_{12}=\ln(1+\xi_{12})
\label{eq:logn2a}
\end{eqnarray}
\n
In the limiting case of large separation $X_{12}\rightarrow 0$ we can write down the two point PDF 
\be
p_{\rm ln}(\delta_1,\delta_2)= p_{\rm ln}(\delta_1)p_{\rm ln}(\delta_2)[1+ b_{\rm ln}(\delta_1)\xi_{12}b_{\rm ln}(\delta_2)]; \quad\quad  b_{\rm ln}(\delta_i)= \Delta_i/\Sigma_{i}.
\ee
\n
It is however easier to estimate the cumulative or integrated bias associated with objects beyond a certain density threshold $\delta_0$.
This is defined as $ b_{\rm ln}(\delta>\delta_0)=\int_{\delta_0}^{\infty} p_{\rm ln}(\delta) b_{\rm ln}(\delta) d\delta / \int_{\delta_0}^{\infty} p_{\rm ln}(\delta) d\delta$.
In the low variance limit ${\bar\xi}_2 \rightarrow 0$ the usual Gaussian result is 
restored $b(\delta)= \delta/{\bar\xi}_2$. The parameters $\Lambda,\Lambda_i, X_{12}, \Sigma$ that 
we have introduced above can be expressed in terms of the two-point (non-linear) correlation function 
$\xi_{12} = \la \delta_1\delta_2 \ra$ and the nonlinear variance $\sigma^2 = \la \delta^2 \ra$ of the smoothed density field.
To understand the construction of lognormal distribution, we introduce a Gaussian PDF in variable $x$;
$p(x) = (2 \pi \Sigma^2)^{1/2} \exp[-(x-\mu)^2/2\Sigma^2] $. With a change of variable $x = \ln(t)$ we can
write down the PDF of $y$ which is a lognormal distribution $p(t) = (2\pi\Sigma^2)^{1/2} \exp [-(\ln(t) -\mu)^2/2\Sigma^2]/t$.
The extra factor of $(1/x)$ stems from the fact: $dt/t = dx$. Note that $y$ is positive definite and 
is often associate with $\rho/\rho_0 = 1 +\delta$ which means $\la t \ra =1$. 
The moment generating function for the lognormal in terms of the mean $\mu$ and the variance $\Sigma$ has the
following form: $\la t^n \ra = \exp(n\mu + n^2 \Sigma^2/2)$.
This however leads to the fact
that if the underlying distribution of $x$ or the density is Gaussian we will have to impose the
condition: $\mu=-\Sigma^2/2$. Here in our notation above $\Sigma$ is the variance of the underlying
Gaussian field. The variance of $t$ defined as $\la t^2 \ra - \la t \ra^2 = \exp(\Sigma^2) -1 = \sigma^2$. So we can write
$\Sigma^2 = \ln(1+\sigma^2)$. This is the result that was used above. The generalization to
two-point or bi-variate PDF can be achieved following the same arguments and can be found in \citep{KTS01}.

In contrast to the lognormal model the widely-used non-local ansatz for the evolution of the variance that
was introduced by \citep{HKLM}, the evolved nonlinear two-point correlation function is
linked to that of the initial or linear two-point correlation at a different scale: 
$\bar\xi_2^{nl}(R_{\rm nl})=f_{nl}[\bar\xi_2^{lin}(R_{lin})]$. The two different length scales are related by the following
expression: $R_{\rm nl}^3= (1+\bar\xi^{nl}_2(R_{\rm nl}))R_{lin}^3$. Such as an ansatz is derived using pair conservation equation.
The non-linear (Eulerian) length scale $R_{\rm nl}$ is linked to the linear (Lagrangian) 
length scale $R_{\rm lin}$ from where the 
structure has collapsed. Numerical simulations are typically used for the determination of 
the fitting function $f_{\rm nl}$  \citep{PD94}. However, simpler asymptotic power-law 
forms exist in different regimes of gravitational clustering \citep{MuPa97}.
%

The validity and limitations of the one-point and two-point PDFs have been 
studied extensively in the literature against N-body simulations. 
In \cite{B92,B94} it was shown that the PDF computed from the perturbation theory
in a weakly nonlinear regime approaches the lognormal distribution function only when
the primordial power spectrum is locally of the form $P(k) \propto k^{n_{e}}$ with the
effective local spectra slope of the power spectrum $n_{e} \sim -1$. It was also
shown that in the weakly nonlinear regime the lognormal distribution is equivalent
to the hierarchical model with a generating function $\cal G(\tau) = \exp(-\tau)$.
This leads to the following skewness and kurtosis parameters: 
\be
S^{\eta}_3 = 3 + \sigma^2; \quad\quad  S^{\eta}_4 = 16 + 15\sigma^2 + 6 \sigma^4 + \sigma^6.
\ee
In general the $\sigma^2 \rightarrow 0$ leads to  $S_p^{\eta} = p^{p-2}$. 
On this basis \cite{BK95} argues that the agreement of lognormal PDF with
numerical simulations should be interpreted as purely accidental and the
success of the lognormal model is simply related to the fact that for all
scales relevant to cosmology the CDM power spectrum can be approximated with
a power law with effective slope $n_{e} \approx -1$. However subsequent 
studies using numerical simulation it was shown by various authors  
 that the lognormal distribution very accurately describes the cosmological
distribution functions even in the nonlinear regime $\sigma \le 4$ for a relatively 
high values of density contrast $\delta < 100$ (see e.g. \cite{KTS01}). 

There is no complete analytical description of gravitational clustering in the highly nonlinear
regime. However several dynamical approximations were proposed in the past to mimic 
certain features of gravitational clustering neyond the weakly nonlinear clsutering.
The  {\it Frozen Flow} Approximation (FFA) approximation proposed by \citep{MLMS92} is one such
approximation. using perturbative techniques it was shown by  \citep{MSS94} that the
FFA develops exactly the same generating function as the lognormal approximation 
in the quasilinear regime.

The error estimates for various lower order $S_p$ (right panel) and the variance (left-panel)
are shown in Figure (\ref{fig:sn_err}). A complete analytical formalism for calculation of error are
given in \cite{VaMuBa05}. It requires the knowledge of higher order $S_p$ parameters.
In our calculation of error the higher order $S_p$ parameters are modelled according to
lognormal distribution. Different levels of noise as chareterized by the parameters that
describe intrinsic ellipticity distribution $\epsilon$ and number density of galaxies/arcmin$^2$ or 
$n_g$ are considered.

%
\section{ The PDF and bias of smoothed redshift-resolved convergence maps} 
\label{sec:pdf_bias}
For computing the probability distribution function of the smoothed 
convergence field for individual tomographic maps $\kappa^{(i)}(\theta_0)$, we will begin by constructing
its associated cumulant generating function for individual tomographic bins $\Phi^{(i)}_{1+\kappa(\theta_0)}(y)$.
The construction is based on modelling of the volume-averaged higher order correlation function $\langle
\kappa_{(i)}^p(\theta_0) \rangle_c$ in terms of the matter correlation hierarchy:
\begin{equation}
\Phi^{(i)}_{1 + {\kappa(\theta_0)}}(y) = y + \sum_{p=2}^ {\infty} {{\langle
\kappa_{(i)}^p(\theta_0) \rangle_c} \over \langle \kappa_{(i)}^2 (\theta_0 )
\rangle_c^{p-1}} y^p.
\end{equation}
\begin{figure}
\begin{center}
{\epsfxsize=6. cm \epsfysize=6. cm {\epsfbox[27 426 315 709]{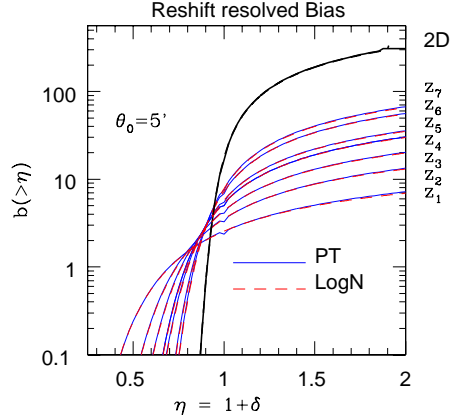}}}
\end{center}
\caption{The cumulative bias $b(>\eta)$ of the {\em reduced} convergence $\eta^{ }=(1+\delta)$ is plotted as a function $\eta$
for various redshift bins. The smoothing angular scale is $\theta_0=5'$. As before two different approximations are 
considered. The lognormal approximation (solid lines) and the perturbative calculations (dashed lines) reproduce 
nearly identical results. The curves that saturates at a higher values of cumulative bias for higher values of $\eta$ correspond
to larger smoothing angular scales. }
\label{fig:eta_bias}
\end{figure}
Now using the expressions for the higher moments of the convergence $\kappa(\theta_0)$
in terms of the matter power spectrum, Eq.\ref{kappa_variance} 
and Eq.\ref{hui} gives:
\begin{equation}
\Phi^{(i)}_{1 + {\kappa({\theta_0})}}(y) \equiv \sum_{p=1} {S^{(i)\kappa}_p \over p!}y^p   = y + \int_0^{r_s} \sum_{p=2}^{\infty}
{ 1 \over p!} S^{\eta}_p{\omega_{(i)}^p(r) \over d_A(r)^{2(p-1)} (r)}
\Big [ {\exnew \over\avi} \Big ]^{ (p-1)} y^p; \quad\quad \bar\xi_2^{(i)\kappa} \equiv \la \kappa^2(\theta_0)\ra_c .
\end{equation}
We can now use the definition of $\phi(y)$ for the matter cumulants to
express $\Phi_{1 + \kappa(\theta_0)}(y)$, in terms of $\phi(y)$:
\begin{equation}
\Phi^{(i)}_{1+\kappa_{\theta_0}}(y) =  \int_0^{r_s} dr
\Big[ { d_A^2(r) \avi \over  \exnew} \Big
] \phi \Big [{\omega_{(i)} (r) \over d_A^2 (r)} {\exnew \over \avi}y \Big ] -y \kappa_{(i)}^{\rm min}; \quad
\kappa^{\rm min}_{(i)} = - \int_0^{r_s}\; dr\;\omega_{(i)}(r) .
\end{equation}
The extra term comes from the $p=1$ term in the expansion 
of $\Phi_{1+\kappa_{\rm  }(\theta_0)}$.
Note that we have used the fully non-linear generating function $\phi$
for the cumulants, though we will use it to construct a generating 
function in the quasi-linear regime. The analysis becomes much easier if we define a new reduced convergence 
field:
\be
\eta_{(i)}({\theta_0}) = {{( \kappa^{\rm min}_{(i)}-\kappa_{(i)}({\theta_0})}) /
\kappa^{\rm min}_{(i)}} = 1 + {\kappa_{(i)}({\theta_0})/ |\kappa^{\rm min}_{(i)}| }.
\ee
Here the minimum value of $\kappa_{(i)}(\theta_0)$ i.e. $\kappa^{\rm min}_{(i)}$ occurs
when the line-of-sight goes through regions that are completely empty of
matter (i.e. $\delta = -1$ all along the line of sight in a redshift window that defines a specific bin i.e. Eq.(\ref{eq:omegai})).
While $\kappa_{(i)}({\theta_0})$ depends on the smoothing
angle, its minimum value $\kappa^{\rm min}_{(i)}$
depends only on the source redshift and background geometry of the
universe and is independent of the smoothing radius. With the reduced
convergence $\eta$, the cumulant generating function is given by,
\begin{equation}
\Phi^{(i)}_{\eta} (y) = { 1 \over [\kminnew]} \int_0^{r_s} dr
\Big [{ d_A^2(r) \over \kminnew}{ \avi \over \exnew }\Big ] \phi \Big [
{\kminnew \over d_A^2(r)} {\exnew \over \avi}y \Big ].
\end{equation}
The thus constructed cumulant generating function  
 $\Phi_{\eta}(y)$ satisfies the normalization constraints $S^{\eta}_1 = S^{\eta}_2 = 1$.
The scaling function associated with $P(\eta)$ can now be 
 easily related with the matter scaling function $h(x)$ 
introduced earlier:
\begin{equation}
h^{(i)}_{\eta} (x) = - \int_{-\infty}^{\infty} { dy \over 2 \pi i} \exp (x
y) \Phi^{(i)}_{\eta} (y); \quad\quad
h_{\eta}^{(i)} (x) = { 1 \over [\kminnew]}
\int_0^{r_s}  dr 
\Big [ {\avi \over  \exnew
\kminnew } \Big ]^2 
   h \Big ({\avi x \over \omega_{(i)}(r) 
\exnew \kminnew } \Big ).  
\end{equation}
\n
While the expressions derived above are exact, and are derived for the most
general case, using only the small angle approximation, they can be
simplified considerably using further approximations. In the following we will
assume that the contribution to the $r$ integrals can be
replaced by an average value coming from the maximum of $\omega_{(i)}(r)$,
i.e. $r_c$ ($0<r_c<r_s$). So we replace $\int f(r) dr$
by $1/2 f(r_c)\Delta_{r}$ where $\Delta_{r}$ is the
interval of integration, and $f(r)$ is the function of comoving radial distance $r$ under
consideration.  Similarly we replace the $\omega(r)$ dependence in 
the ${\bf k}$ integrals by $\omega_{(i)}(r_c)$. 
\begin{equation}
|\kappa_{(i)}^{\rm min}| \approx {1\over 2} r_s \omega_{(i)}(r_c), 
\quad\quad |\bar\xi^{(i)\kappa}|  \approx {1\over 2} r_s  {\omega_{(i)}(r_c)\over d_A(r_c)}{\omega_{(i)}(r_c)\over d_A(r_c)} \Big [ \int {d^2 \bl \over
(2\pi)^2} {\rm P_{\delta}(k)} W_{\rm TH}^2(l\theta_0) \Big ].
\label{eq:approx}
\end{equation}
Under these approximations we can write: $\Phi^{(i)}_{\eta}(y) = \phi^{(i)}(y)$  and  $h^{(i)}_{\eta}(x) = h(x)$.
Thus we find that the statistics of the underlying field $\eta=1+\delta$ and the 
statistics of 
the reduced convergence $\eta$ are exactly the same under this
approximation. Though we derived the results from considering a 
specific form of hierarchical ansatz, the final result is remarkably
general. It simply means that independent of detailed modelling
the reduced convergence will always follow the statistics of underlying
mass density distribution.

Finally, we can express the relations connecting the probability distribution
function for the smoothed convergence statistics
$\kappa_{\rm  }({\theta_0})$, the reduced
convergence $\eta({\theta_0})$, i.e. for individual bins we can write:
\be
p_{(i)}(\kappa) ={p(\eta^{(i)})/|\kappa^{\rm min}_{(i)}|}.
\label{eq:pdf}
\ee
\begin{figure}
\begin{center}
{\epsfxsize=12. cm \epsfysize=6. cm {\epsfbox[27 426 585 709]{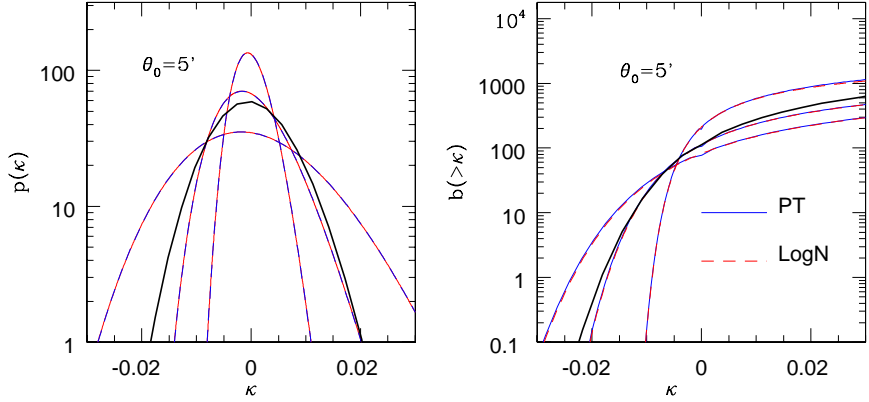}}}
\end{center}
\caption{The left panel shows the PDF of the redshift resolved convergence and the right panel shows the associated cumulative
bias. The smoothing angular scales considered is $\theta_0 = 5'$. Only three tomographic bins are chosen for display to avoid cluttering. 
Two different approximations are used; the lognormal distribution and the perturbative calculations. The approximations 
give near identical results. Three different redshift bins are displayed $z_s=0.698,1.095,1.493$  (top,middle and bottom curve
respectively).}
\label{fig:kappa_pdf_bias}
\end{figure}
This is one of the most important results in this paper. 
A few comments are in order, it is possible to integrate the exact expressions of the scaling functions, 
there is some uncertainty involved in the actual determination of these functions and 
associated parameters such as $\omega, k_a, x_{\star}$ from N-body simulations and its 
unclear how much is there to gain by doing exact calculations that involve approximate picture for
the underlying mass distribution;  see e.g. Munshi et al. (1999), Valageas et al. (1999) and  
Colombi et al. (1996) for a detailed description of the effect  of the finite volume correction involved in their
estimation. Throughout our analysis we have used a top-hat filter for smoothing
the convergence field, but the results can easily be extended to
compensated or Gaussian filters \citep{BV00}. Using  Eq.(\ref{eq:pdf}) one can
derive an approximate expression for the lower order moments which can, in turn, be used
to derive order of magnitude relations $S^{\kappa}_{\rm p} = S^{\eta}_{\rm p}/[k^{\rm min}_{(i)}]^{{\rm p}-2}$ for $\rm p>2$.
The parameters $S^{\eta}_{\rm p}$ for the underlying density contrast $\delta$ are specified 
by the choice of a specific hierarchical model. The computation of error bars for these lower order 
moments can be done using formalism developed in \cite{MuCo03}.

In Figure (\ref{fig:eta_pdf}) we show the PDF of the reduced convergence $\eta$ for smoothing angular scale $\theta_0=5'$ for the lognormal and hierarchical approximation discussed above,
for individual redshift bins as well as for a projected survey. 
\subsection{The bias associated with convergence maps} 
To compute the bias associated with the peaks in the smoothed convergence $\kappa$ field we
have to first develop an analytic expression for the generating field
$\beta_{\kappa}(y_1, y_2)$ for the convergence field $\kappa^{\rm  }(\theta_0)$. We will
avoid displaying the smoothing angular scale $\theta_0$ for brevity. Throughout the statistics are
for smoothed convergence fields. For that we will use the usual definition for the two-point 
cumulant correlator $C_{pq}$ for the convergence field (for a complete
treatment see Munshi \& Coles, 1999b):
\be
C^{(ij)}_{pq} = {\langle \kappa_{(i)}(\oh_1)^p \kappa_{(j)}(\oh_2)^q \rangle_c /  [{\bar\xi_2^{(i)}}]^{p-1}[{\bar\xi_2^{(j)}}]^{q-1}
 \xi^{(ij)}_{12} }.
\ee
We will show that, like its density field counterpart the
two-point generating function for the convergence field $\kappa_{\rm  }$ 
can be
expressed (under certain simplifying assumptions) as a product 
of two one-point generating functions $\beta^{\rm  }(y)$
which can then be directly related to the bias associated with
``hot-spots''in the convergence field. 
\begin{equation}
\beta^{(ij)}_{\eta}(y_1,y_2) =  \sum_{p,q}^{\infty} {C^{\eta(ij)}_{pq} \over p! q!} y_1^p y_2^q = 
\sum_{p}^{\infty} {C^{\eta(i1)}_{p1} \over p!} y_1^p \sum_{q}^{\infty} {C^{\eta(j1)}_{q1}
\over q!} y_2^q  = \beta^{(i)}_{\eta}(y_1) \beta^{(j)}_{\eta}(y_2).
\end{equation}
\n
It is clear that the factorization of generating function 
depends on the factorization property of the cumulant correlators i.e.
$C^{\eta}_{pq} = C^{\eta}_{p1} C^{\eta}_{q1}$. Note that such a factorization is 
possible when the correlation of two patches in the directions 
$\oh_1$ and $\oh_2$ $\two$  is smaller compared to the variance
$\one$ for the smoothed patches
We will now use the integral expression for cumulant correlators
(Munshi \& Coles 1999a) to
express the generating function which in turn uses the hierarchical 
{\em ansatz} and the far field approximation as explained above
\be
\beta^{(ij)}_{\kappa}(y_1, y_2) = \sum_{p,q}^{\infty} {C^{\eta(ij)}_{pq} \over p! q! } { 1 \over 
[\bar\xi_{2}^{(i)}]^{p-1}}{ 1 \over 
[\bar\xi_{2}^{(j)}]^{q-1}} { 1 \over \xi^{12}_{ij}}
\int_0^{r_s} dr
{ \omega_{(i)}^{p}(r) \omega_{(j)}^q(r) \over d_A(r)^{2(p+q -1)} }[{\cal J}_{\theta_{12}}(r)]
 [\var]^{p+q-2} y_1^p y_2^q.
\ee
\n
It is possible to further simplify the above expression by separating the
summation over dummy variables $p$ and $q$, which will be useful to
establish the factorization property of the two-point generating function
for bias $\beta^{(ij)}(y_1, y_2)$.
%
We can now decompose the double sum over the two indices into two
separate sums over individual indices. Finally, using the definition of
the one-point generating function for the cumulant correlators 
we can write: 
\be
\beta^{(ij)}_{\kappa}(y_1, y_2) = \inc dr\;  d_A^2(r) {\corr \over \two } { \onei \over \var }{ \onej \over \var } \nonumber  
\beta^{(i)}_{\eta} \Big ( { y_1 \over \onei} {\omega_{(i)}(r) \over d^2_A(r)
} \var  \Big )  \beta^{(j)}_{\eta} \Big ( { y_2 \over \onej} {\omega_{(j)}(r) \over d^2_A(r)
} \var  \Big ).
\ee
\begin{figure}
\begin{center}
{\epsfxsize=12. cm \epsfysize=6. cm {\epsfbox[27 426 584 709]{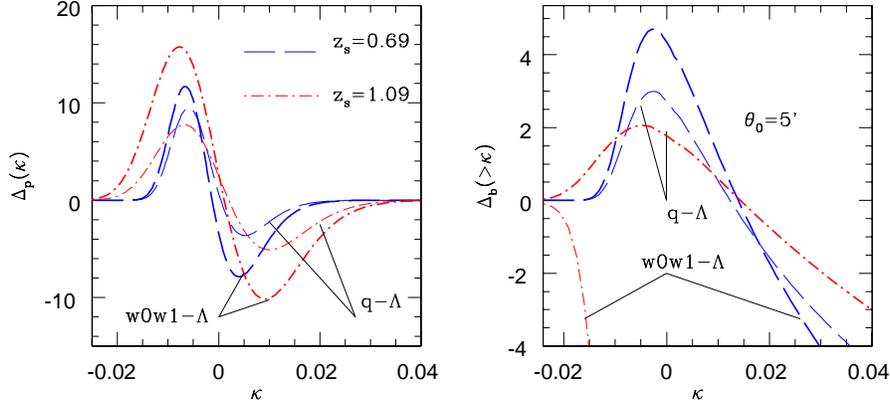}}}
\end{center}
\caption{We compare the PDF and cumulative bias associated with tomographic convergence maps for various tomographic bins
for two different dark energy models against the $\Lambda$CDM model. The left panel plots the ratio 
$\Delta_p(\kappa)= p(\kappa)-p_{\Lambda{\rm CDM}}(\kappa)$ for a smoothing angular scale of $\theta_0=5'$. The right panel
depicts the ratio $\Delta^{(i)}_b(>\kappa)= b^{(i)}(>\kappa)-b^{(i)}_{\Lambda{\rm CDM}}(>\kappa)$ for the cumulative bias.
Three different redshift bins are displayed $z_s=0.698,1.095$. For a given smoothing angular scale and 
a fixed redshift, the (thick) curves with higher positive peak heights
correspond to the model $w_0=-1, w_1=1$ and the (thin) ones with lower peak heights correspond to $w_0=-0.9$, $w_1=0$.}
\label{fig:dark}
\end{figure}
\n
The above expression is quite general within the small 
approximation and large separation approximations, and is valid for any
given specific model for the generating function ${\cal G}_{\delta}(\tau)$. However it is easy
to notice that the projection effects as encoded in the line of sight
integration do not allow us to write down the two-point generating
function $\beta_{\kappa}(y_1, y_2)$ simply as a product of two one-point generating functions $\beta_{\eta}(y)$ as
was the case for the density field $1+ \delta$.  
\be
\beta_{\kappa}^{(ij)}(y_1, y_2) = \inc \; dr\; {d_A(r)\over [\kappa^{min}_{(i)}]}{d_A(r)\over [\kappa^{min}_{(j)}]}  
{\corr \over \two } { \one^2 \over \var^2 
} 
\beta^{(i)}_{\eta} \Big ( { y_1 \over \onei} {\omega_{(i)}(r) \over d^2_A(r)
} {\var \over |\kminnew|} \Big ) 
\beta^{(j)}_{\eta} \Big ( { y_2 \over \onej} {\omega_{(j)}(r) \over d^2_A(r)
} {\var \over |\kminnew|} \Big ).
\ee
We use the follwoing equation in assocition with  Eq.(\ref{eq:approx}) to simplify the above expression:
\be
\two  \approx {1\over 2} r_s {\omega_{(i)}(r_c) \over d_A(r_c) } {\omega_{(j)}(r_c) \over d_A(r_c) }
\Big [ \int {d^2 \bl \over (2\pi)^2} {\rm P_{\delta}(k)} W_{\rm TH}^2(l \theta_0) \exp [il \cdot \theta_{12}] \Big ].
\ee
\n
Use of these approximations gives us the leading order contributions
to these integrals and we can check that to this order we recover the 
factorization property of the generating function i.e. $\beta^{(ij)}_{\eta}(y_1,
y_2) = \beta^{(i)}_{\eta}(y_1) \beta^{(j)}_{\eta}(y_2) = 
\beta^{(i)}_{1+\delta}(y_1) \beta^{(j)}_{1+\delta}(y_2)$.
So it is clear that at this 
level of approximation, due to the factorization property of the cumulant 
correlators, the bias function $b_{\eta}(x)$ associated with the peaks
in the convergence field $\kappa_{\rm  }$,  beyond certain threshold, possesses a 
similar factorization property too as its
density field counterpart. Earlier studies have established such a 
correspondence between convergence field and density field 
in the case of one-point probability distribution function $p(\delta)$
(Munshi \& Jain 1999b),
\begin{equation}
b_{\eta}^{(i)}(x_1) h_{\eta}^{(i)}(x_1) b_{\eta}^{(i)} (x_2) h_{\eta}^{(j)} (x_2) = 
 b^{(i)}_{1 + \delta}(x_1) h^{(i)}_{1 + \delta}(x_1) b^{(j)}_{1+\delta} (x_2) h^{(j)}_{1+\delta} (x_2),
\end{equation}
\n
where we have used the following relation between $\beta_{\eta}(y)$
and $b_{\eta}(x)$. For all practical purpose we found that the differential bias 
as defined above is more difficult to measure from numerical
simulations as compared to its 
integral counterpart where we concentrate on the bias associated with
peaks above certain threshold. The cumulative bias $b_{\eta}(>x)$
can also be defined in an analogus manner:
\begin{equation}
b_{\eta}(x) h_{\eta}(x) = -{ 1 \over 2 \pi i}
\int_{-i\infty}^{i\infty} dy \tau (y) \exp (xy); \quad
b_{\eta}(>x) h_{\eta}(>x) = -{ 1 \over 2 \pi i}
\int_{-i\infty}^{i\infty} dy {\tau (y)\over y} \exp (xy).
\label{eq:bias1}
\end{equation}
\n
It is important to notice that, although the bias $b(x)$
associated with the convergence field $\kappa_{\rm  }$ and the underlying density
field are identical, the variance associated with the density field is
very high, while projection effects substantially reduce the variance in the convergence field.
This indicates that we have to use
the integral definition of bias to recover it from its generating
function; see Eq.(\ref{eq:bias1}). 
Now writing down the full two point probability distribution function 
for two correlated spots in terms of the convergence field $\kappa_{\rm  }(\theta_0)$ and
its reduced version $\eta$:
\ben
\quad\quad\quad\quad\quad\quad\quad\quad\quad\quad\quad\quad\quad\quad\quad\quad p^{(ij)}(\kappa^{\rm  }_1,\kappa^{\rm  }_2)d\kappa^{\rm  }_1 d\kappa^{\rm  }_2 = p^{(i)}(\kappa^{\rm  }_1) p^{(j)}(\kappa^{\rm  }_2)( 1
+ b^{(i)}(\kappa^{\rm  }_1) \xi^{(ij)}_{12} b^{(j)}(\kappa_2)) d\kappa^{\rm  }_1 d\kappa^{\rm  }_2, \\ 
\quad\quad\quad\quad\quad\quad p_{\eta}(\eta_1, \eta_2)d\eta_1 d\eta_2 = p^{(i)}_{\eta}(\eta_1) p^{(j)}_{\eta}(\eta_2)( 1
+ b^{(i)}_{\eta}(\eta_1) \xi^{\eta}_{12} b^{(j)}_{\eta}(\eta_2)) d\eta_1 d\eta_2. 
\een
\n
Using Eq.(\ref{eq:pdf}) that $p^{(i)}(\kappa)
= {p_{\eta}(\eta)/{|\kappa^{(i)}_{\rm min}}|}$ we also notice that $\xi^{(ij)}_{12} = 
{\xi^{\eta}_{12}/{[\kappa^{(i)}_{\rm min}]}{[\kappa^{(j)}_{\rm min}]}}$,
from which we can now write: 
\be
b_{(i)}(\kappa) = {b^{(i)}_{\eta}(\eta)/|{\kappa^{(i)}_{\rm min}}|}.
\label{eq:bias}
\ee
Together with Eq.(\ref{eq:pdf}), Eq.(\ref{eq:bias}) can be used to construct analytical estimates
of pdf and bias about individual bins. In addition these results are applicable to the modelling
of joint PDFs involving two separate redshift bins. 

Figure (\ref{fig:eta_bias}) shows the bias associated with the reduced convergence for individual bins
as well as for the entire survey. The smoothing angular scale is $\theta_0=5'$. In Figure (\ref{fig:kappa_pdf_bias})
shows the PDF and bias associated with the convergence $\kappa$. In Figure (\ref{fig:dark}) we plot the
difference in PDF between various cosmological scenarios. 
\section{Effect of Noise on one- and two-point PDF}
\begin{figure}
\begin{center}
{\epsfxsize=12.5 cm \epsfysize=6. cm {\epsfbox[25 434 585 709]{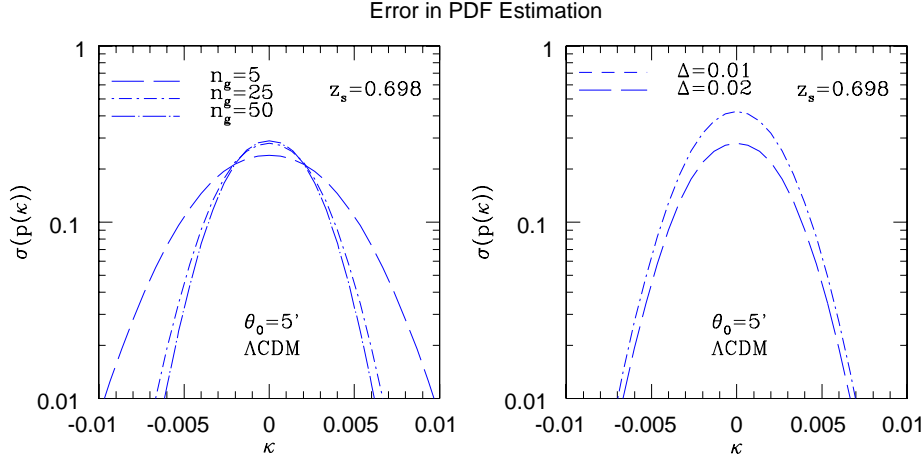}}}
\end{center}
\caption{The sacatter in estimation of binned PDF for a given intrinsic ellipticity distribution and sky coverage
is displayed. We assume an all-sky coverage. The effect of intrinsic ellipticity distribution is displayed 
in the left panel. The right panel depicts the effect of binning. The angular scale in each case is $\theta_0=5'$
and the redshift is $z_s=0.698$. The expression for $\sigma(p(\kappa))$ is given in Eq.(\ref{eq:s2n}). For the 
left panel we consider $\Delta=0.02$ and for the right panel $n_g=25$. A background $\Lambda$CDM cosmology is 
accumed for these calculations.}
\label{fig:s2n1}
\end{figure}

\begin{figure}
\begin{center}
{\epsfxsize=6. cm \epsfysize=6. cm {\epsfbox[27 426 315 709]{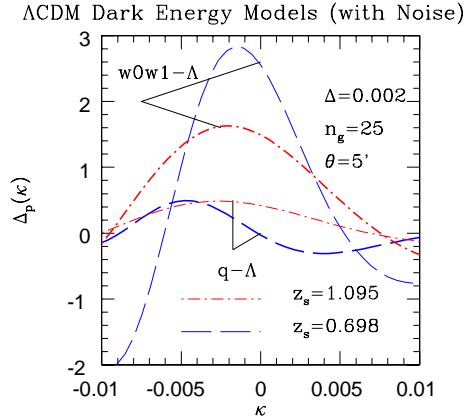}}}
\end{center}
\caption{The difference of noisy $\Lambda$CDM PDF and dark energy models $\Delta_p(\kappa)=p(\kappa)-p_{\Lambda \rm CDM}(\kappa)$ 
is plotted as a function of $\kappa$. The smoothing angular scale, bin size and galaxy number density is as depicted.
The scatter in estimation is smaller compared to the difference in the PDFs considered. The comsological parameters
considered are the same as the ones in Figure-(\ref{fig:dark}). The two survey configurations that we have considered
both produces near identical results.}
\label{fig:10}
\end{figure}

The PDF we have considered so far are free from noise.
In this section we will present the results of estimates of error relating to the PDFs, those at the level 
of one-point and two-point PDFs. These results will generalise the ones found by \cite{MuCo03}
for lower order moments and later by \cite{VaMuBa05}. Inclusion of noise can be incorporated 
through a convolution. We will assume the noise to be Gaussian and uncorrelated with the signal.
However the variance of the noise will depend on the surface density of galaxies in individual bins.
With these simplifying assumption, for the i-th tomographic bin we can write: 
\be 
p^{(i)}_n(\kappa) = \int_{-\infty}^{\infty}\;p^{(i)}(\kappa-n)\;p^{(i)}_G(n)\;dn.
\ee
Here $p^{(i)}_G(n)$ is the noise PDF assumed Gaussian, and $p^{(i)}_n(\kappa)$ is the convergence PDF 
in the presence of noise (the subscript $G$ denotes Gaussian). We take $\sigma^2_{\kappa} = \sigma^2_{\epsilon}/(2 n_g \pi\theta_0^2)$.
Here $\sigma_{\epsilon}$ is intrinsic ellipticity distribution of galaxies and $n_g$ is the number denisty of galaxies and
$\theta_0$ is the smoothing angular scale.
The two-point PDF can also be modified to include the effect of noise in a similar manner.
The equivalent expression for 2PDF can be written as:
\be
p^{(ij)}_n(\kappa_1,\kappa_2) = p^{(i)}_n(\kappa_1)p^{(j)}_n(\kappa_2)( 1 + b^{(i)}_n(\kappa_1)\xi^{ij}_{12}b^{(j)}_n(\kappa_2)),
\label{eq:2pt1}
\ee
Which is obtained by convolving the noise PDF with the 2PDF:
\be
p^{(ij)}_n(\kappa_1,\kappa_2) = \int_{-\infty}^{\infty}\;p^{(ij)}(\kappa_1-n_1,\kappa_2-n_2)\; p^{(i)}_G(n_1)\; p^{(j)}_G(n_2)\;dn_1\;dn_2.
\label{eq:2pt2}
\ee
Comparing Eq.(\ref{eq:2pt1}) and Eq.(\ref{eq:2pt2}) we can write the expression for the noisy bias $b_n(\kappa)$ as:
\be
b^{(i)}_n(\kappa) =  \int_{-\infty}^{\infty} p^{(i)}(\kappa-n)b^{(i)}(\kappa-n)p^{(i)}_G(n)dn \; / \;\int_{-\infty}^{\infty} p^{(i)}(\kappa-n)p^{(i)}_G(n)\;dn.
\ee
Notice that depending on redshift distribution of sources, the noise maps $n^{(i)}$ and $n^{(j)}$ can be different for two tomographic bins. We also
assumed that noise in different bins are statistically independent. The cumulative bias for the i-th tomographic bin that include noise $b^{(i)}_n(>\kappa)$ 
can be expressed in terms of $p^{(i)}_n(\kappa)$  $b^{(i)}_n(\kappa)$ just as its noise-free counterpart: 
\be
b^{(i)}_n(>\kappa)= \int_{\kappa}^{\infty} p^{(i)}_n(\kappa) b^{(i)}_n(\kappa) dk / \int_{\kappa}^{\infty} p^{(i)}_n(\kappa) dk.
\ee 
Errors associated with binned tomographic noisy PDF can be analysed using following quantites:
\be
{\rm N} = n_g \pi \theta_0^2 = 314 \left (  {n_g \over 100 \;{\rm arcmin}^{-2}}\right ) \left ( \theta_0 \over 1 \;{\rm arcmin}  \right )^2
\ee
Here $n_g$ is the number density of galaxies, $\theta_s$ is the smoothing angular scale in arc-minutes for a given survey strategy.
For a given survey we denote the area covered by $A$ and introduce a parameter $N_c$ which will be used in expressing the 
signal-to-noise estimates of the PDF $p(\kappa)$. We define the following variable that will be useful in qunatifying scatter in a noisy PDF: 
\be
{\rm N}_c = {A \over (2\theta_0)^2} =  2.7 \times 10^4 \left( {A \over 300 \;{\rm degree}^2} \right ) \left ( {\theta_0 \over 1 \;{\rm arcmin}} \right )^{-2}.
\ee
Finally the scatter $\sigma(p(\kappa))$ in the measured convergence power sepctra $p(\kappa)$ can be expressed as \citep{VaMuBa05}:
\be
{\sigma(p^{(i)}_n(\kappa)) \over p^{(i)}_n(\kappa)}  = \left [ {1 \over {\rm N}_c} \left ( {1 \over 2 p_n^{(i)}(\kappa) \Delta } -1 \right ) \right ]^{1/2} 
\label{eq:s2n}
\ee
These expression can be modified and used to compute the scatter in individual redshift bins by simply changing $\bar n_g$ to
surface density of individual bins $\bar n_g^{(i)}$ and $p(\kappa)$ to $p^{(i)}(\kappa)$. The source density for individual
bins for a given survey can be computed using $\bar n^{(i)} = \int_{z_i}^{z_{i+1}} p_s(z) dz$. The bin width $\Delta$ is left as a
free parameter. In Figure (\ref{fig:s2n1}) we have plotted the scatter $\sigma(p^{(i)}_n(\kappa)$ as a
function of intrinsic ellipticity
distribution and bin width for a smoothing angular scale of $\theta_0 =5'$ and redshift $z_s=0.698$. 
The results for the difference in noisy PDFs are plotted in Figure (\ref{fig:10}).
\section{Conclusions}
\label{sec:conclu}
%
%
Previous tomographic studies of weak lensing have typically worked with 
the lower order cumulants; we have generalized here these
results to the case of the entire one- and two-point PDF, which contain 
information about the cumulants to an arbitrary order.
The construction was performed using a generating function formalism
based on hierarchical ansatz and a lognormal model. Our analysis generalizes previously
obtained results derived for projected or 2D surveys. Though 
we have considered a top-hat filer convergence maps due to their simplicity, similar results
can be obtained for related statistics such as the shear components or aperture mass $M_{ap}$ \citep{BV00}.
%
%

The PDFs for the individual bins are constructed by generalization of the 
previously introduced global variable $\kappa^{\rm min}$, for individual bins i.e.$\kappa^{\rm min}$,  
that was used in the context of 2D projected maps.
Next, using $\kappa_{(i)}^{\rm min}$,  reduced variable $\eta^{(i)}$ is defined for 
each individual bins whose statistics can directly be linked 
to that of underlying density contrast $\delta$. The convergence in individual bins
can then be mapped to unique values of $\eta=1+\delta$ for a given smoothing angular scales $\theta_0$.
%

For modelling the statistics of underlying density contrast $\delta$ we have assumed two
completely different model: the hierarchical ansatz along with its perturbative 
counterpart as well as the lognormal distribution. Both these approximations
have been used successfully in various cosmological contexts. There are a wide class of models that
are available under the general category of hierarchical ansatz. 
The main motivation behind our choice of a specific hierarchy is simplicity.
In recent years more sophisticated models of hierarchical clustering have been
proposed which rely more on inputs from numerical simulations. The generic results
we have derived here indeed can be improved using such modelling though the fundamental
characteristics will remain unchanged.
%
In our treatment we find, in agreement with \cite{MuWa03}, the dynamical and geometrical 
contribution can be treated separately. The geometrical effects are completely encoded in
a parameter $\kappa^{min}$. The reduced convergence as defined is independent of 
the background geometry of the universe and essentially probe the evolution of
gravitational clustering. We showed that a set of $\kappa^{\rm min}_{(i)}$ defined
for a given set of redshift slices are adequate to characterize not only
individual PDFs for each bin but it is also sufficient to study the 
joint two-point PDF among two different bins. The PDF of the reduced convergence
$\eta^{(i)}$ for individual bins or joint PDFs for a pair of bins 
generalizes the earlier studies where the projected or 2D maps 
were considered in a straight forward manner. 
%

We also note that the construction of convergence maps is difficult compared to the direct evaluation of 
non-Gaussian statistics from shear maps. On the other hand convergence statistics 
can directly be modelled at arbitrary order whereas for shear field the computation
is done mostly order by order manner. An independent analysis of convergence 
maps constructed from shear maps should therefore be useful in constraining
various errors which might get introduced during various stages of data reduction.
%

In our analysis we have ignored the noise from intrinsic ellipticity distribution of galaxies
as well as from shot noise resulting from finite number of galaxies that are used
to trace the underlying mass distribution. These issues have been dealt with 
in great detail in \cite{MuCo03,VaBaMu04}. Dividing the source population into bins reduced the
number-density of sources. This in turn will increase the level of noise or the
scatter in the estimator. In our analysis we have considered two different survey configurations,
i.e. LSST and DES and found that for our choice of tomographic bins the one- and two-point 
PDFs are very similar in nature. 
%

The lognormal distribution has already been used to model 
the statistics of weak lensing observables \citep{Mu00,TTHF02} and the clustering of Lyman alpha
absorption systems e.g. \citep{BD97}. One-to-one mapping of initial density fields to
evolved density fields using maps that are consistent with lognormal distribution
function was not found to be very successful and the success of a lognormal distribution
function in reproducing the statistics of gravitational clustering still remains 
somewhat unclear. 
%

Tomographic weak lensing surveys can be cross-correlated with external data sets
including frequency cleaned maps of secondaries from  ongoing CMB surveys;
e.g. the thermal Sunyaev-Zeldovic (tSZ) maps or $y$-maps that will be available from
CMB surveys such as Planck. The cross-correlation with tomographic
information can help to understand the evolution of cosmological pressure fluctuations
responsible for tSZ effect with redshift. The formalism presented here is 
perfectly suitable for such analysis. Detailed results of such analysis will be presented
elsewhere. In addition to the weak lensing surveys the Supernova pencil beam surveys 
might also benefit for the results presented here.
%

To summarize, we have extended results derived in three different previous papers
\citep{MuCo03,VaMuBa05,VaBaMu04} to take into
account tomographic bins within which the photometric redshift are available. The results
obtained previously for one-point PDF are now extended to two-point PDF.
These results can provide an alternative to usual Fisher-matrix analysis 
that is employed to optimize survey strategies. We have concentrated mainly on 
analytical results in this paper. The numerical results regarding optimization of survey strategy using 
these results will be considered elsewhere.

\section{Acknowledgements}
\label{acknow}
DM and PC acknowledges support
from STFC standard grant ST/G002231/1 at School of Physics and
Astronomy at Cardiff University where this work was completed. 
We would like to thank Alan Heavens, Patrick Valageas, Ludo van Waerbeke and Sanaz Vafei for many useful discussions.
The numerical results were obtained using a modified version of a code made available to us
for computing the PDF and bias by Francis Bernardeau. 
\bibliography{paper.bbl}
\appendix
\end{document}